\documentclass[letterpaper,11pt,twocolumn]{article}

\usepackage{array,amssymb,amsmath}
\usepackage{graphics,graphicx}
\usepackage{natbib}
\usepackage{aas_macros}

\begin{document}
\pagestyle{plain}
\pagenumbering{arabic}

\title{Polarization of prompt and afterglow emission of \\ Gamma-Ray Bursts}

\author{Stefano Covino$^{1}$ and Diego G\"otz$^{2}$ \\
\\
$^{1}$ INAF / Brera Astronomical Observatory, Via Bianchi 46, 23907, Merate (LC), Italy\\
$^{2}$ AIM--CEA/DRF/Irfu/Service d'Astrophysique, Orme des Merisiers, 91191 Gif-sur-Yvette, France
}

\date{}
\maketitle




\abstract{
Gamma-ray bursts and their afterglows are thought to be produced by an ultra-relativistic jet. One of the most important open questions is the outflow composition: the energy may be carried out from the central source either as kinetic energy (of baryons and/or pairs), or in electromagnetic form (Poynting flux). While the total observable flux may be indistinguishable in both cases, its polarization properties are expected to differ markedly. The prompt emission and afterglow polarization are also a powerful diagnostic of the jet geometry. Again, with subtle and hardly detectable differences in the output flux, we have distinct polarization predictions. In this review we briefly describe the theoretical scenarios that have been developed following the observations, and the now large observational datasets that for the prompt and the afterglow phases are available. Possible implications of polarimetric measurements for quantum gravity theory testing are discussed, and future perspectives for the field briefly mentioned.

}


\section{Introduction}

Polarimetric measurements can provide useful complementary information about the physical processes at work in Gamma-Ray Bursts (GRBs). Indeed several different possible scenarios have been invoked to interpret the large amount of observational data now available for the GRBs. Most of the theoretical efforts have been applied to the so-called ``standard model'' \citep{reesstdmod,mesresstmmod,revpir} that, although a fully satisfactory picture is still missing, offers the best (while not unique) interpretative scenario for the polarimetric observations. 

In this review we have separated the GRB phenomenology in the two traditional phases: prompt  and afterglow, mainly due to the different observational techniques. Further subdivisions (plateau phase, steep decay, etc.) are mentioned when required. In Sections \ref{sec:prompt_theory} and \ref{sec:the} we summarize the current status of theoretical modeling of polarization in prompt and afterglow observations, in Sections \ref{sec:prompt_obs} and \ref{sec:aftobs} we present the current status of the observations, and in Section \ref{sec:liv} we provide some insights into the implication of these measurements for fundamental physics. Some general conclusions are finally drawn in Section\,\ref{sec:conclusions}.

\section{Polarization in the prompt phase}

\subsection{Theory}
\label{sec:prompt_theory}
The expected level of polarization of the prompt $\gamma$-ray emission in GRBs has been estimated by several authors for different models, or variations within them. In most cases, the observed $\gamma$-ray emission is due to the synchrotron radiation from relativistic electrons.
To have a high radiative efficiency and to allow for the short time scale variability in the GRB light curves, these electrons have to be in the fast cooling regime. Their time-averaged distribution is a broken power law, $n(\gamma)\propto \gamma^{-p'}$ with $p'=p+1$ above $\Gamma_\mathrm{m}$ and $p'=2$ below, where $\Gamma_\mathrm{m}$ is the minimum Lorentz factor of the injected distribution of electrons, and $p\simeq 2-2.5$ its slope \citep{sari98}. The intrinsic polarization level of the synchrotron radiation, $\Pi_\mathrm{syn}=(p'+1)/(p'+7/3)$ \citep{ryb} is then of the order of $\Pi_\mathrm{syn}=(p+2)/(p+10/3)\simeq 75\%$ above $\nu_\mathrm{m}$ and $\Pi_\mathrm{syn}=9/12\simeq 70 \%$ below, where $\nu_\mathrm{m}$, the peak of the spectrum in $\nu F_\nu$, is the synchrotron frequency of electrons at $\Gamma_\mathrm{m}$. High polarization levels can also be reached if inverse Compton scatterings are the dominant radiative process.\\ 

Actually different scenarios in terms of radiation processes and observer's viewing angle can be envisaged to explain the presence of polarized emission during the prompt phase of GRB emission. They can be roughly divided in two families: intrinsic models and geometric models, for which peculiar observing conditions are required.

\begin{enumerate}

\item \textit{Synchrotron emission from shock-accelerated electrons in a relativistic jet with an ordered magnetic field contained in the plane perpendicular to the jet expansion}. This scenario is compatible with the magnetic field being carried by the outflow from the central source, as the poloidal component decreases much faster with radius than the toroidal one. 
The polarization level at the peak of a given pulse can be as high as $\Pi/\Pi_\mathrm{syn}\sim 0.8$, i.e. $\Pi\sim 60 \%$, leading to a maximum time-averaged polarization in long intervals of $\Pi/\Pi_\mathrm{max}\sim 0.6 $, i.e. $\Pi\sim 45 \%$ in this case \citep{granot03a,granot03b,nakar03}. 
The main requirement for this model to apply is to have a uniform magnetic field in space, i.e. with a coherence spatial scale $R \theta_{B}$ with $\theta_{B} \gtrsim 1/\Gamma$. In this scenario the polarization level and angle can vary during the burst only if the magnetic field is not  
uniform in time, while the opposite needs to be true (i.e. a magnetic field constant in time) to explain a high level of the time-integrated polarization \citep{nakar03}. But a magnetic field anchored in the central engine and carried by the outflow to large distance \citep[see e.g.][]{spruit01} is not the only possibility. A magnetic field generated at the shock could also work in principle, and even favour variability, but this requires a process capable of locally increasing the field coherence scale (the field is most probably initially generated on small, skin-depth, scales). The existence of such a process is not yet settled in our present knowledge of the micro-physics in mildly relativistic shocks. 
Note that the condition $\theta_{B} \gtrsim 1/\Gamma$ is really necessary only in the pulses with the highest level of polarization. If $\theta_{B}$ is smaller, so that a number $N\sim (\Gamma\theta_{B})^{-2}$ of mutually incoherent patches are present in the visible region, the level of polarization will decrease, but the variability (both of the polarization level and angle) will increase \citep{granot03a}. Indeed if the radiating electrons are accelerated in internal shocks \citep{rees94,kobayashi97,daigne98}, the Lorentz factor associated with the individual shells is necessarily varying in the outflow, which can be an additional source of variability for the polarization. If $\theta_\mathrm{B}$ and $1/\Gamma$ are close, the number of coherent patches in the visible region could vary from one pulse to another. 
This scenario could hence produce time variable polarization, as long as the coherence scale $\theta_\mathrm{B}$ of the field is larger than $1/\Gamma$ in most of the emitting regions. 
A potential difficulty remains: an additional random component of the magnetic field is probably necessary to allow for particle acceleration in shocks. This component would
reduce the coherence of the field and hence the level of polarization by some factor, which  is however difficult to estimate, as the intensity of this additional component is not well constrained \citep{granot03a,nakar03}; \\


\item \textit{Synchrotron emission from a purely electromagnetic outflow}. In this scenario the GRB is powered by the rotational energy of a magnetar-like progenitor \citep[e.g.][]{metzger11}. It is first converted into magnetic energy by the dynamo action of the unipolar inductor, propagated in the form of Poynting-flux-dominated flow, and then dissipated at large distances from the source. The estimated level of polarization in this case
 is comparable with the previous scenario (up to $\sim$50\%) \citep{lyutikov03}. In addition, a magnetic field with a large coherence scale is naturally expected in such a purely electromagnetic outflow. One potential difficulty is, however, related to the mechanism responsible for the energy dissipation. In this scenario, the energy has to be extracted from the magnetic field before being radiated. Therefore magnetic dissipation must occur in the emitting region, changing the field geometry, which becomes  probably much less ordered, reducing the final level of polarization by a large factor \citep{lyutikov03,nakar03}. This effect is however difficult to estimate, as the details of the physical processes that could lead to
magnetic dissipation in such an outflow are still far from being understood. In addition in this scenario there would not be a natural explanation of polarization level or angle variability. \\

\item \textit{Synchrotron emission from shock-accelerated electrons in a relativistic jet with a random field generated at the shock and contained in the plane perpendicular to the jet velocity}. A high level of polarization can be obtained even with a random magnetic field of the jet is observed from just outside its edge  \citep{ghisellini99,waxman03}.
The polarization at the peak of a given pulse can reach $\Pi/\Pi_\mathrm{syn}\simeq 0.8$, i.e. $\Pi \simeq 60 \%$ resulting in a time-integrated value of the order of $\Pi/\Pi_\mathrm{syn}\simeq 0.5-0.6$, i.e. $\Pi\simeq 40-45\%$ \citep{granot03a,granot03b,nakar03}. However these high values are obtained if the jet is seen with $\theta_\mathrm{obs}\simeq \theta_\mathrm{j}+1/\Gamma$, where $\theta_\mathrm{j}$ is the opening angle of the jet and $\theta_\mathrm{obs}$ the angle between the line-of-sight and the jet axis. Such viewing conditions are rare, except if $\theta_\mathrm{j}\sim 1/\Gamma$. 
Variability of the polarization level is expected if the Lorentz factor is varying in the outflow, as for instance in the internal shock model. 
Observations are made at $\theta_\mathrm{obs}=\theta_\mathrm{j}+k/\Gamma$ with $k$ being larger for emitting regions with a larger Lorentz factor. The maximum level of polarization is obtained for $k\sim 1$ whereas the flux decreases with $k$ for $k\ge 0$. On average, the highest polarization should therefore not be found in the brightest pulses. This is however difficult to test, as the intrinsic luminosity of each pulse is not necessarily the same. In addition, as the emission from several pulses can be superposed, 
the measured polarization level, which is flux-weighted, could be reduced by a sizeable factor \citep{granot03b}.
Finally, the observed polarization can also be reduced if the jet edges are not sharp enough \citep{nakar03}. 
As shown by \citet{granot03b}, similar conditions as for scenario (3) would be required for a scenario where the field is ordered but parallel to the jet, leading to the same conclusions.\\

\item \textit{Synchrotron emission from shock-accelerated electrons in a relativistic jet with an ordered magnetic field parallel to the jet velocity}. This case has been studied by \citet{granot03b} and gives very similar results to model (3). The viewing conditions have to be the same and it suffers from the same difficulties as listed above. \\

\item \textit{Inverse Compton emission from relativistic electrons in a jet propagating within a photon field ("Compton drag" model)}.
Inverse Compton scattering of external light by the electrons in highly relativistic narrowly collimated jets as the origin of GRBs has been suggested for the first time by \citet{shaviv95}. The level of polarization in this scenario can be even higher than for the synchrotron radiation and reach $60-100\%$, but only under the condition that the jet is narrow with $\Gamma\theta_{j}\lesssim 5$  \citep{shaviv95,lazzati04}. The maximum level of polarization is again obtained for $\theta_{obs}\simeq \theta_{j}+1/\Gamma$. These viewing conditions are very similar to those of model (3). This scenario predicts a lower level of polarization for the afterglow phase \citep[see later, ][]{dadopol}.
Again, the polarization is reduced if the edges of the jet are not sharp enough. Variability of the Lorentz factor will again result in a varying polarization, with the same difficulties regarding the final level of polarization as in model (3). However, variations of the Lorentz factor could possibly be less large in this scenario as part of the variability of the light curve can be related to the inhomogeneity of the ambient photon field.\\

\item Independently from the emission process (synchrotron  or inverse Compton), \textit{fragmented fireballs} (shotguns, cannonballs, sub-jets) can produce highly polarized emission, with a variable polarization amplitude. The fragments are responsible for the single pulses and have different intrinsic properties (such as Lorentz factors), opening angles, orientations with respect to the observers and magnetic domains. \citep[e.g.][]{lazzati09}. In this case
the most polarized pulses are those which have about one tenth of the flux of the main pulse, i.e. an anti-correlation between the polarization level
an the GRB pulse flux is expected.

\end{enumerate}


\subsection{Observations}
\label{sec:prompt_obs}

The measurement of polarization during the prompt phase of GRBs has always been challenging. This is mainly due to the fact that no wide field gamma-ray polarimeter with a large effective area has yet been flown, and that many of the measurements attempted to date have been performed with instruments which have some polarimetric capabilities, but do not have an explicit polarimetric 
oriented design. In addition at odds to the afterglow emission, the GRB prompt emission is very limited in time, mostly less than $\sim$100 s, and hence in spite of the high average flux of GRBs, the total number of collected photons is often too limited to derive statistically stringent limits for polarization.

\subsubsection{Early Results}
The first attempt to  measure linear polarization in the prompt emission of GRBs was reported by \citet{coburn03}. They used the Reuven Ramaty High Energy Solar Spectroscopic Imager (RHESSI) observations of GRB 021206. RHESSI has an array of nine large-volume (300 cm$^{3}$) coaxial germanium (Ge) detectors with high spectral resolution, and has been designed to study solar flares in the 3 keV--17 MeV energy range. In the soft gamma-ray energy range (0.15--2.0 MeV) the dominant photon interaction in RHESSI is Compton scattering. Polarization at high energies can be measured, thanks to the polarization dependency of the differential cross section for Compton scattering 
\begin{equation}
\frac{d\sigma}{d\Omega} = \frac{r_{0}^{2}}{2}\left(\frac{E^{\prime}}{E_{0}}\right)^{2}\left(\frac{E^{\prime}}{E_{0}}+\frac{E_{0}}{E^{\prime}}-2 \sin^{2}\theta \cos^{2}\phi \right)
\label{eq:cross}
\end{equation}
where $r_{0}^{2}$ is the classical electron radius, $E_{0}$ the energy of the incident photon, $E^{\prime}$
the energy of the scattered photon, $\theta$ the scattering angle, and $\phi$ the azimuthal angle relative
to the polarization direction. Linearly polarized photons scatter preferentially perpendicularly to the incident
polarization vector. Hence by examining the angles of scattering of the photons among the Ge detectors, one can in principle derive the degree and angle of linear polarization of the incident photons.

\citet{coburn03} reported a high level of linear polarization of $\Pi$=80$\pm$20\% (close to and beyond the theoretical value, see Section $\ref{sec:prompt_theory}$) at a high level of confidence ($>$ 5.7$\sigma$) for GRB 021206. In the RHESSI detector a small fraction of the incident photons undergoes a Compton scattering in a given detector before being photoelectrically absorbed in a second detector (or undergo other scatterings). The accurate analysis of the photon scattering angles can be exploited to measure the degree of polarization of the incident photons. In addition RHESSI is a rotating instrument (4 s period), which presents the advantage of averaging out the effects of asymmetries in the detector and the passive materials. 
GRB 021206 was a quite bright GRB with a fluence of 1.6$\times$10$^{-4}$ erg cm$^{-2}$ in the 25--100 keV energy band, and a peak flux of 2.9$\times$10$^{-5}$ erg cm$^{-2}$ s$^{-1}$. Scattered photons represent about 10\% of the total events. \citet{coburn03} interpreted the angular modulation measured in the data as a high-level polarization signal.  

However subsequent re-analyses of the same data set could not confirm this result reporting a polarization level compatible with zero \citep{rutledge04,wigger04}. These authors show that the number of suitable events for polarization analysis has been over-estimated by a factor 10 (830$\pm$150 versus 9840$\pm$96), since spurious coincidences had been counted as Compton scattering events, implying in the end statistics too small to be able to measure any polarization signal, even for a 100\% polarized source. Despite the non-confirmation of the RHESSI result, the former work had the merit of triggering some theoretical work about the possibility of polarized emission associated with the prompt phase of GRBs (see Section \ref{sec:prompt_theory}).

Another early attempt to measure the polarization level of the prompt emission of GRBs has been performed by \citet{willis05}, who used the data from the Burst and Transient Sources Experiment (BATSE) on board the Compton Gamma-Ray Observatory (CGRO). They studied two GRBs, 930131 and 960924, and, by modelling the scattering of the gamma-ray photons by the Earth atmosphere, they reported evidence of high level of polarization in both bursts, $\Pi >$35\% and $\Pi >$50\%, respectively. But unfortunately this method did not allow the authors to statistically constrain these results, but called for further independent confirmations to ascertain whether the prompt emission of GRBs is highly polarized or not.

\subsubsection{IBIS and SPI on board INTEGRAL}

At the time of its discovery by the \textit{INTEGRAL} Burst Alert System (IBAS) \citep{ibas}, GRB 041219A \citep{mcbreen06,gotz11} was among the top 1\% in terms of GRB fluence. This prompted different attempts to measure its polarization with the instruments that observed it. The first attempt was performed using the SPI spectrometer on board \textit{INTEGRAL}. SPI \citep{spi} is made by individual hexagonal Ge detector and the measuring technique used is similar to the one used for RHESSI. \citet{kalemci07} reported a high level of polarization for this GRB ($\Pi=98\pm33\%$), but could not constrain the systematics of their measurements. Using the same dataset but a more sophisticated analysis technique, where the multiple event scatterings in the SPI spectrometer have been compared to a GEANT4 Monte Carlo simulation predicted response to a polarized source flux, \citet{mcglynn07} were able to measure the degree of linear polarization over the brightest pulse of the GRB (lasting 66 s) to $\Pi=63^{+31}_{-30}\%$ and the polarization angle to $P.A.=70^{+14}_{11}$ degrees. However, they could not completely exclude the presence of a systematic effect mimicking the observed polarization degree. 

GRB041219A was also observed by the Imager on Board the INTEGRAL Satellite \cite[IBIS;][]{ibis}. Thanks to its two superposed pixellated detection planes -- ISGRI \citep{isgri}, made of CdTe crystals (4$\times$4$\times$2 mm) and operating in the 15 keV--1 MeV energy range, and PICsIT, made of CsI bars (9$\times$9$\times$30 mm) and operating in the 200 keV--10 MeV energy range\citep{picsit}, IBIS can be used as a Compton Polarimeter. By examining the scatter angle distribution of the detected photons in the two planes

\begin{equation}
N(\phi)=S[1+a_{0}\cos 2(\phi-\phi_{0})],
\label{eq:azimuth}
\end{equation}

one can derive the polarization angle, $PA =  \phi_{0} - \pi /2 + n \pi$, and the polarization fraction 
$\Pi= a_{0}/a_{100}$, where $a_{100}$ is the amplitude expected for a 100\% polarized source derived
by Monte Carlo simulations \citep[see][]{forot08}. IBIS has been used to measure the polarization from bright gamma-ray sources such as the Crab nebula \cite{forot08}, and the black hole binary Cyg X--1 \cite{laurent11}. 

\begin{figure*}[ht!]
\centering
\includegraphics[width=15cm]{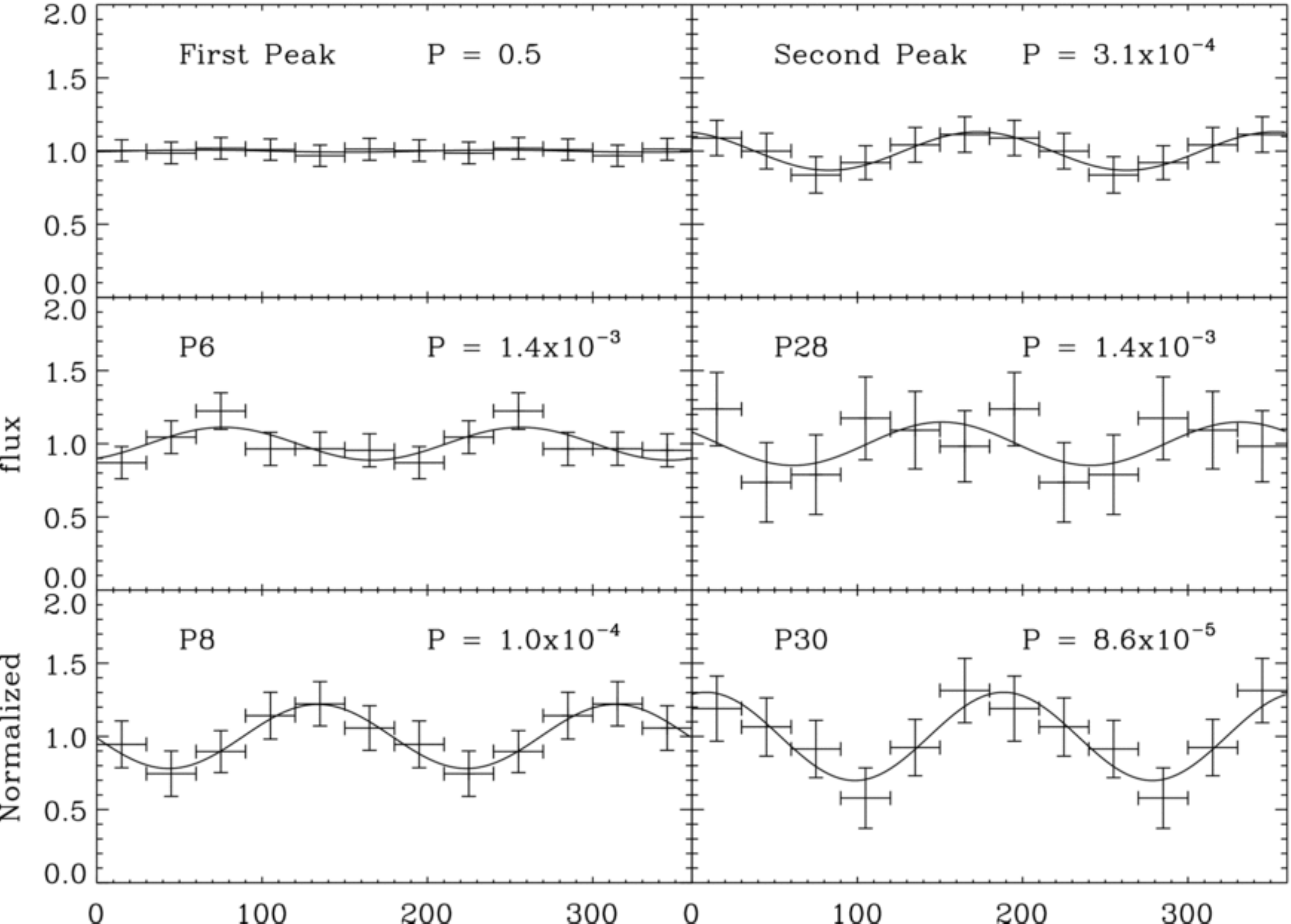}
\caption{Polarigrams of the different time intervals that have been analysed for GRB041291A (see Table \ref{tab:pola}). For comparison purposes, the curves have been normalized to their average flux level. The crosses represent the data points (replicated once for clarity) and the continuous line the fit done using Eq. \ref{eq:azimuth}. For each polarigram the probability, $P$, is shown that the polarigram measured corresponds to an un-polarized ($\Pi<$1\%) source. From \citet{gotz09}}.
      \label{fig:lc}
\end{figure*}

Indeed to perform the polarization analysis, the source flux as a
function of $\phi$ is derived, and the scattered photons are then divided in 6
bins of 30$^{\circ}$. To improve the signal-to-noise ratio in each bin, one can take
advantage of the $\pi$-symmetry of the differential cross section,
i.e. the first bin contains the photons with
$0^{\circ}<\phi<30^{\circ}$ and $180^{\circ}<\phi<210^{\circ}$, etc.
The chance coincidences (i.e. photons interacting in both detectors within a time window of 3.8 $\mu$s
but not related to a Compton event), have been estimated using the data before the GRB and subtracted from each detector image following the procedure described in \citet{forot08}. The derived detector images are then deconvolved to obtain sky images, where the flux of the source in each bin is measured by fitting the instrumental PSF to the source peak, building a so-called polarigram of the source, see Fig. \ref{fig:lc}.

The polarigrams can then be fitted with Eq. \ref{eq:azimuth} using a least
squares technique to derive $a_{0}$ and $\phi_{0}$. 
Confidence intervals on $a_{0}$ and $\phi_{0}$ cannot, on the other hand, be derived from the fit, since
the two variables are not independent. They were derived from the
probability density distribution of measuring $a$ and $\phi$ from $N$
independent data points over a $\pi$ period, based on Gaussian
distributions for the orthogonal Stokes components (see Eq. 2 in
\citealt{forot08}).

\begin{figure*}[ht!]
\centering
\includegraphics[width=8cm]{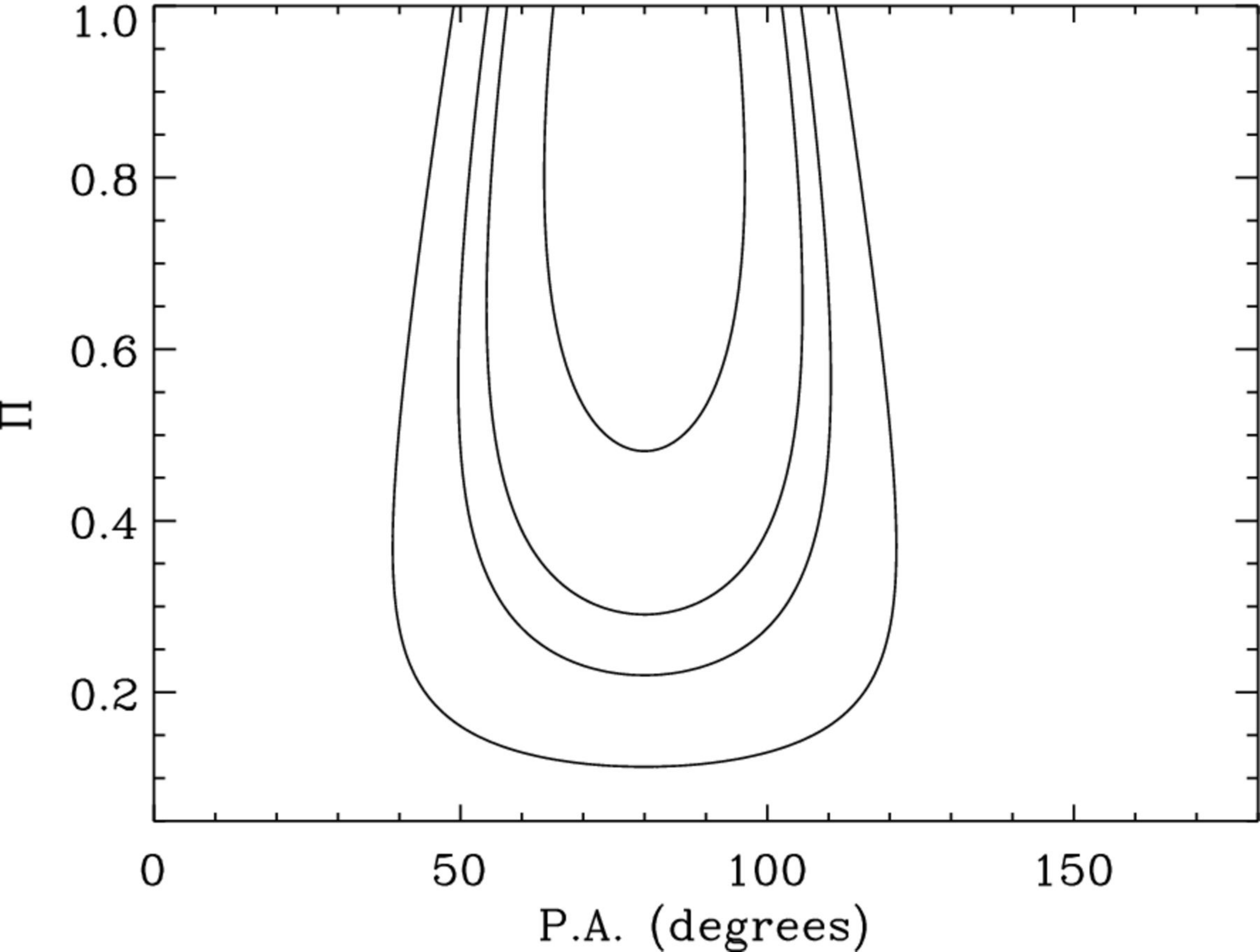}
\includegraphics[width=8cm]{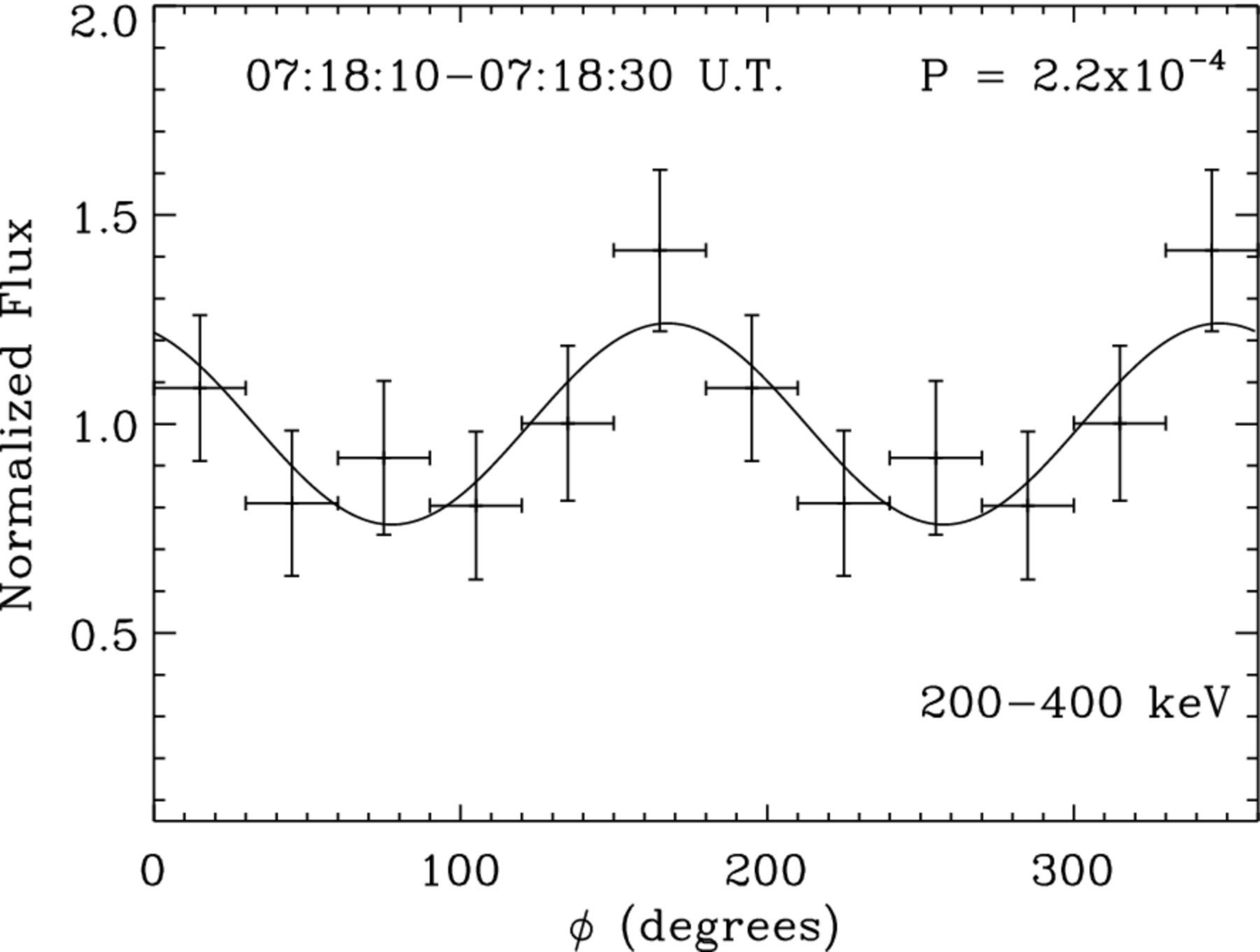}
\caption{Left: The 68, 90, 95 and 99\% confidence contours for the $\Pi$ and PA parameters. Right:  Polarigram of GRB 140206A in the 200--400 keV energy band. The crosses represent the data points (replicated once for clarity) and the continuous line the fit done on the first six points using equation \ref{eq:azimuth}. The chance probability P of a non-polarized ($<$1 \%) signal is also reported. The normalized flux corresponds to N($\phi$)/S.. From \citet{gotz13}}.
      \label{fig:140206A}
\end{figure*}

\begin{table*}[ht]
\caption{Polarization results for the different time intervals. From \cite{gotz09}}
\begin{center}
\begin{tabular}{cccccc}
\hline\hline
Name & T$_{start}$ & T$_{stop}$ & $\Pi$ & $PA$ & Image\\
& U.T.  & U.T. & \% & degrees & SNR \\
\hline
First Peak & 01:46:22 & 01:47:40 & $<$4 & -- & 32.0\\
Second Peak & 01:48:12 & 01:48:52 & 43$\pm$25 & 38$\pm$16 & 20.0\\
P6 & 01:46:47 & 01:46:57 & 22$\pm$ 13& 121$\pm$17 & 21.5\\
P8 & 01:46:57 & 01:27:07 &65$\pm$26 & 88$\pm$12 & 15.9\\ 
P9 & 01:47:02 & 01:47:12 &61$\pm$25 & 105$\pm$18 & 18.2\\
P28 & 01:48:37 & 01:48:47 &42$\pm$42 & 106$\pm$37 & 9.9\\
P30 & 01:48:47 & 01:48:57&90$\pm$36 & 54$\pm$11 & 11.8\\
\hline
\end{tabular}
\end{center}
Errors are given at 1 $\sigma$ c.l. for one parameter of interest.
\label{tab:pola}
\end{table*}

Using the same method polarization could be measured for two other GRBs (061122\footnote{A consistent polarization measurement has been obtained with SPI by \cite{mcglynn09}.} and 140206A, see Fig. \ref{fig:140206A}) with IBIS \citep{gotz13,gotz14}, see Tab. \ref{tab:polasummary}, but no time-resolved analysis could be performed
due to the limited statistics, making GRB041219A the only GRB for which a time variable polarization signal could be measured to date with IBIS.

\begin{table*}[ht!]
\centering
\caption{Summary of recent GRB polarization measurement by IBIS/SPI and GAP.} 
\label{tab:polasummary}
\begin{tabular}{ccccccc}
\hline
GRB & $\Pi$   & Peak energy & Fluence & Energy Range  & Redshift & Instrument\\ 
& (68\% c.l.) & (keV) & (erg cm$^{-2}$) &  & $z$ &\\
\hline
041291A & 65$\pm$26\% & 201$^{+80}_{-41}$ & 2.5$\times 10^{-4}$ & 20--200 keV & 0.31$^{+0.54}_{-0.26}$ & IBIS, SPI\\ 
06122 & $>$60\% & 188$\pm$17 & 2.0$\times 10^{-5}$ & 20--200 keV & 1.33$^{+0.77}_{-0.76}$ & IBIS, SPI\\ 
100826A & 27$\pm$11\%& 606$^{+134}_{-109}$ & 3.0$\times 10^{-4}$ & 20 keV--10 MeV & 0.71--6.84$^{1}$ & GAP\\
110301A & 70$\pm$22\% & 107$\pm$2 & 3.6$\times 10^{-5}$ & 10 keV--1 MeV & 0.21--1.09$^{1}$ & GAP\\
110721 & 84$^{+16}_{-28}$\% & 393$^{+199}_{-104}$ & 3.5 $\times 10^{-4}$ & 10 keV--1 MeV & 0.45--3.12$^{1}$ & GAP\\
140206A & $>$48\% & 98$\pm$17 & 2.0$\times 10^{-5}$ & 15--350 keV & 2.739$\pm$0.001 & IBIS\\ 
\hline
\end{tabular}
\\$^{1}$ redshift based on empirical prompt emission correlations, not on afterglow observations.
\end{table*}

\subsubsection{GAP}

High levels of linear polarization could be measured also for three GRBs (100826A, 110301A and 110721) by the Gamma-Ray Burst Polarimeter \citep[GAP;][]{gap} experiment on board the IKAROS spacecraft \citep{yonetoku11,yonetoku12}. GAP is designed to measure the degree of linear polarization in the prompt emission of GRBs in the energy range 70--300 keV. Also in the GAP case the detection principle is the anisotropy of the differential Klein-Nishina cross section for Compton scattering. The GAP consists of a dodecagon (twelve-sided polygon) plastic scintillator with a single non-position sensitive photomultiplier tube of 17 cm in diameter and 6 cm in thickness surrounded by 12 CsI(Tl) scintillators with 5 mm in thickness. The central plastic scintillator serves as a Compton photon scatterer and the angular distribution of scattered photons coinciding in time with the plastic scintillator is measured by the surrounding CsI scintillators each with an angular resolution of 30$^{\circ}$. In fact, by examining the coincidences within a time window of 5 $\mu$s, one can measure an asymmetry in the detector number counts for the CsI detectors.

\begin{figure*}[ht!]
\centering
\includegraphics[width=8cm]{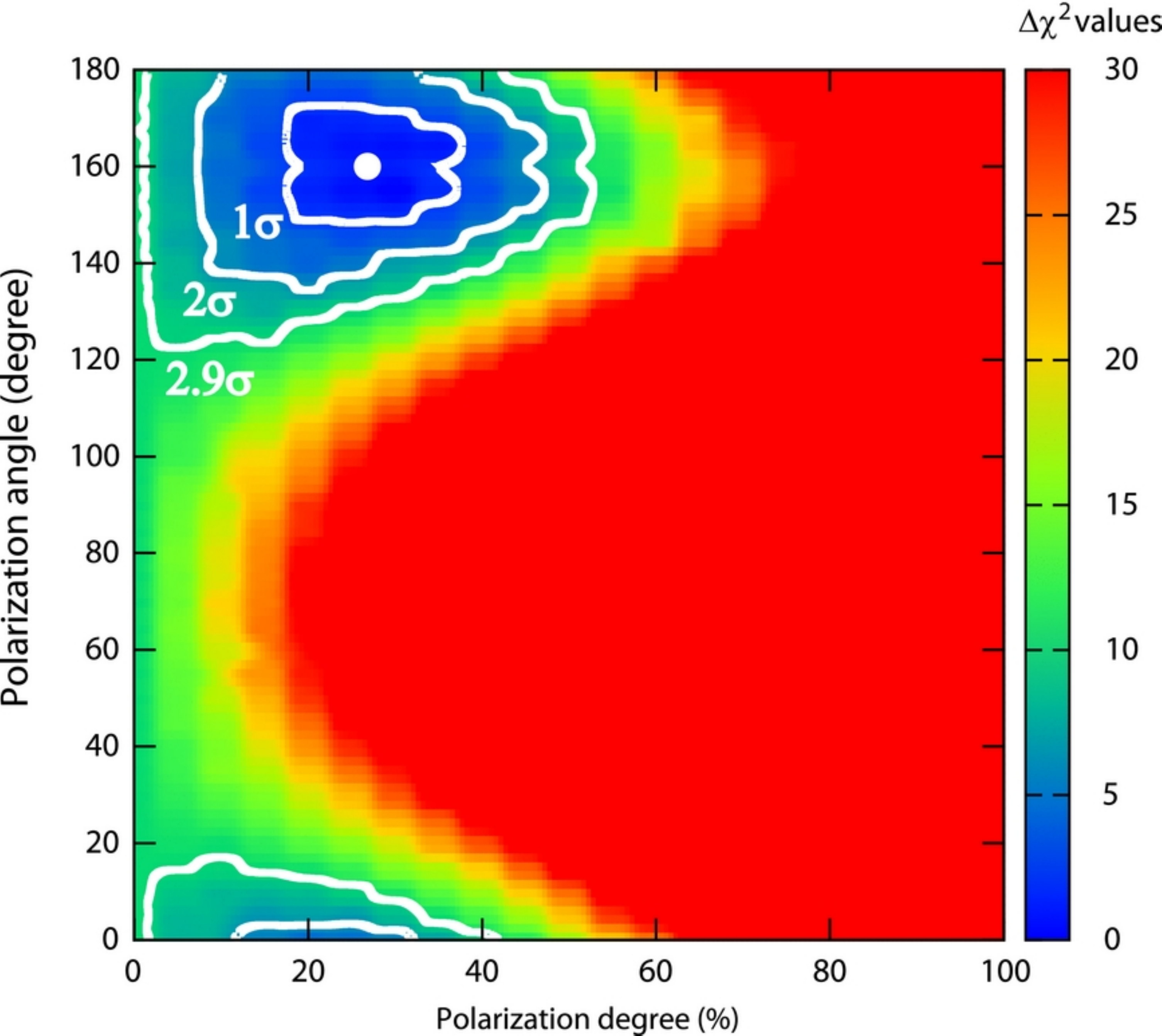}
\includegraphics[width=7.5cm]{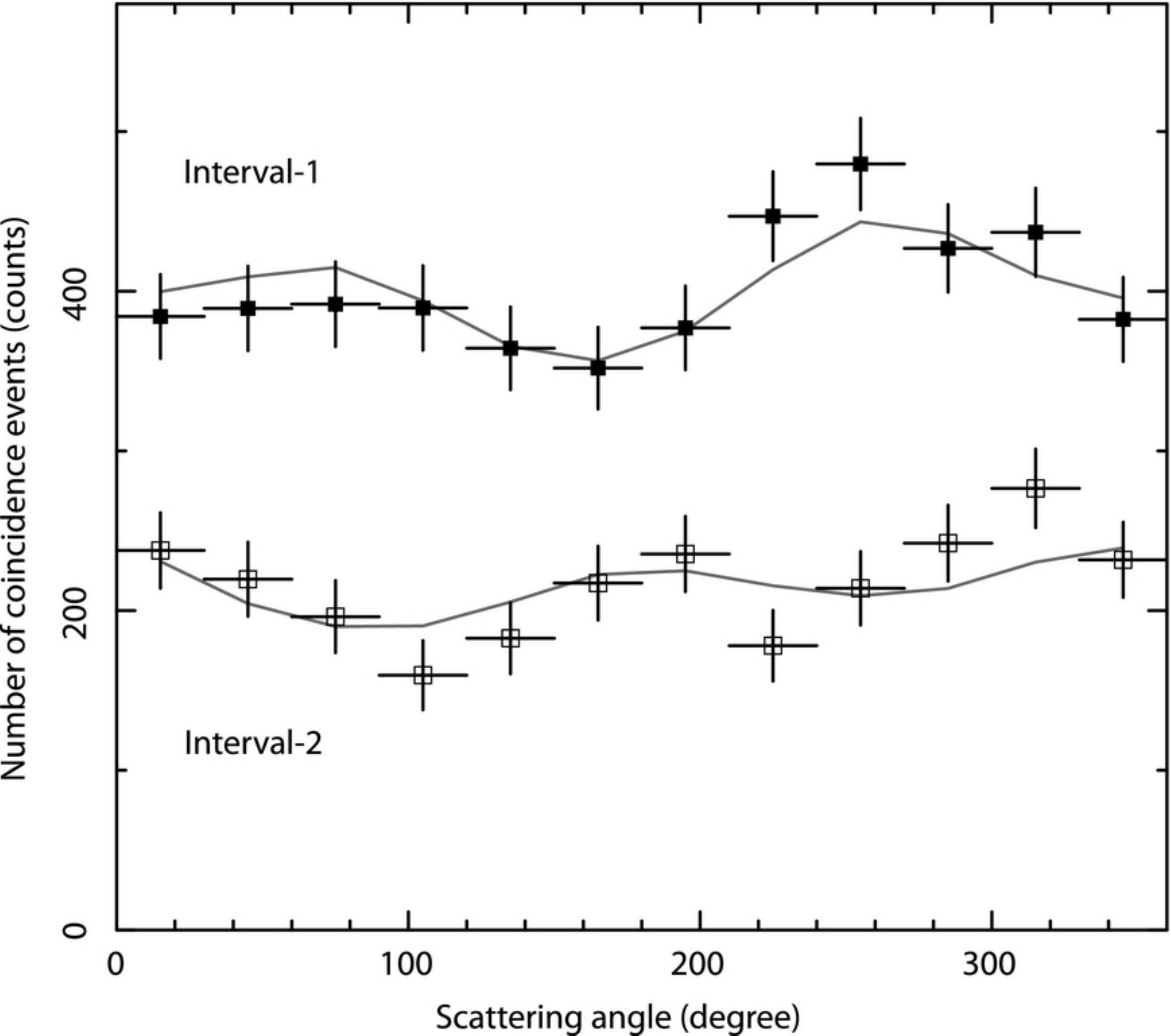}
\caption{Left:$\Delta\chi^{2}$ map of confidence contours in the $\Pi$, $\phi_{p}$) plane for GRB 100826A, obtained by the combined fit of the Interval-1 and -2 data. Here $\phi_{p}$ is the phase angle for Interval-1. The white dot is the best-fit result, and we calculate $\Delta\chi^{2}$ values relative to this point. A color scale bar along the right side of the contour shows the levels of the $\Delta\chi^{2}$ values. The null hypothesis (zero polarization degree) can be ruled out with 99.4\% (2.9$\sigma$) confidence level. Right:  Number of coincidence $\gamma$-ray photons (polarization signals) against the scattering angle of GRB 100826A measured by the GAP in 70--300 keV band. Black filled and open squares are the angular distributions of Compton scattered $\gamma$-rays of Interval-1 and -2, respectively. The gray solid lines are the best-fit models calculated with our GEANT4 Monte Carlo simulations. From \citep{yonetoku11}}.
      \label{fig:100826A}
\end{figure*}

As shown in Table \ref{tab:polasummary} the GAP succeeded to measure the linear polarization for three GRBs. In particular for GRB 100826A (see Fig. \ref{fig:100826A}), which had a similarly high fluence as 041219A, \citet{yonetoku11} were able to measure a change in the polarization angle by dividing the GRB in two $\sim$50 s long time intervals: the angle changed from 159$\pm$18$^{\circ}$ to 75$\pm$20$^{\circ}$ (1 $\sigma$ c.l. for two parameters of interest) with a significance of 3.5 $\sigma$ for the change. For this burst the averaged background coincidence rate is 5.6 counts s$^{-1}$ CsI$^{-1}$, and the total coincidence $\gamma$-rays suitable for polarization analysis are 4281 and 2733 for the first and second intervals respectively. The detector response to a  polarized source has been calculated with GEANT4 Monte Carlo simulations and the predicted modulation curves have been computed for different geometrical and spectral input parameters. The observed modulation curves have then been fitted using a least-squares method to the modelled curves.

\section{Polarization in the afterglow}

Attempts to review the whole subject or some selected topic have been carried out by several authors \citep{bjpol,lazproc,covearlyrev,malesrev,lazrev,lazxrayrev,covxrayrev,kobayashi}. We now discuss at first some of the main theoretical scenarios that have been developed in the context of the ``standard model", and later follow in some detail the observations so far carried out and how they have been modeled in this context. We devote our attention to afterglow phase although mention of phenomena possibly more related to the prompt emission, e.g. X-ray flares or optical emission during the prompt phase itself, are possible. On the contrary, phenomena definitely of high interest but not directly related to the GRB emissions, e.g. polarization of supernovae (SN) associated with GRBs, are not discussed here \citep[see][]{snpol}.

\subsection{Theory}
\label{sec:the}

The concept of an afterglow, following the main, high-energy, GRB emission was probably first explicitly introduced in \citet{PaRh93}. As for all phenomena involving particle acceleration, polarimetry \citep{polbook} is naturally  considered a powerful diagnostic tool. 

The first attempt to derive predictions to be compared with observations came probably by \citet{micropol}. The original idea is indeed still of some interest, and it is based on the observation that a cosmological GRB should appear on the sky as a narrow expanding emission ring \citep[e.g.,][]{ring}. After about a day, the ring radius, $\sim 3\times10^{16}$\,cm\,(t/day)$^{5/8}$, should be comparable to the Einstein radius of a solar mass lens at cosmological distance. Microlensing by an intervening star can therefore significantly affect both the light curve and the polarization signal \citep{microcoll}. The predictions are clearly dependent on the specific afterglow model and outflow energy structure, and on the mechanisms producing the polarized flux. The idea of observing microlensing events for GRBs was originally introduced well before the detection of the first afterglow \citep{micro1,micro2}, and the probability for a stellar microlensing of a source at a cosmological redshift is estimated to be $\sim 0.1 \Omega_* b^2$ \citep{micro3,micro4}, where $\Omega_*$ is the mean density of stellar-mass objects in the universe, in units of the critical density, and $b$ is the impact parameters in units of the Einstein radius. Adopting typical parameters as in \citet{micropol} the probability turns out to be close to unity and the lensing duration is about one day, driven by the emitting area expansion rate.

The afterglow polarization in \citet{micropol} is generated locally at the afterglow emission region, which is modeled as a finite set of discrete patches, each having a coherent and independent magnetic field. The synchrotron radiation emitted by each patch, if the electron energy distribution follows a power-law with index $p$, is polarized at a level \citep{ryb}:
\begin{equation} 
\Pi = \frac{p+1}{p+7/3},
\label{eq:synchroval}
\end{equation}
which for $p \sim 2$ turns out to be $\Pi \sim 0.7$. Clearly, a microlensing phenomenon able to magnify part of the emitting region might allow us to study in detail its magnetic and energy structure. The total polarization observable from the whole afterglow emission depends on the sum of the polarized flux from each patch, which can have random orientation.  In this case the total average polarized flux, $<P> \propto 1/N$, and tends to zero for a large number of independent patches. Given the statistical nature of the sum involved in the derivation of the total polarization, one can expect random variations of both the total polarized flux and position angle in time, during the afterglow evolution. However, in case a lens can magnify a part of the emitting region this is going to dominate the sum and substantially modify the expected observable polarization and offer a powerful diagnostic tool for the magnetic field and energy structure of the afterglow outflow.

The scenario with polarization generated by a large number of independent magnetic domains was further developed by \citet{patchy}. The authors observed that the magnetic fields must be generated in the blast wave because to match the afterglow observations magnetic fields much larger than those typically existent in shock-compressed interstellar medium [ISM, $B \sim \Gamma B_{\rm ISM} \sim 10^{-4} (\Gamma/10^2)$\,G, where $\Gamma$ is the shock Lorentz factor] are required. However, the resulting polarization for an unresolved source depends also on the coherence length of the generated field. If their length grows at about the speed of light, and it is therefore comparable to the thickness of the blast wave, a maximum polarization at about 10\% is expected. The emitting region will be covered by a hundred mutually incoherent patches. The degree and direction of polarization should depend on time. The polarization coherence time is $\sim \epsilon t_0$ with a polarization degree $\Pi \sim 10\epsilon^{3/2}$\%, where $t_0$ is the observing time and $\epsilon (< 1)$ is rate of growth of the coherence length in units of the speed of light. Polarization at a much lower level would imply that the magnetic fields generated at the shock are highly tangled and confined to the shock front.

How magnetic fields can be generated in ultra relativistic shocks is still far from being fully understood \citep{shockmag}. Beyond shock-compression of the ISM magnetic field, it is possible that a magnetic field already existent in any GRB progenitor is carried by the outflow plasma or by a precursor wind. Because of the flux freezing, the field amplitude would decrease as the wind expands and even in the case of a progenitor with very strong magnetic field ($B \sim 10^{16}$\,G) at $R\sim 10^{16}$\,cm the field amplitude would be a few orders of magnitude too low to match the observations. Several authors \citep[e.g.,][]{weibml,inouermi} proposed that relativistic two-stream instabilities can generate magnetic fields with $10^{-5} - 10^{-1}$ of the equipartition energy density ($U_B/8\pi$) in collisionless shocks \citep[see however][]{patchybis}. The generated fields are parallel to the shock front and fluctuate on the very short scale of the plasma skin depth. Since the afterglow synchrotron radiation is beamed toward the observer within a very small opening angle, $\Theta \sim \Gamma^{-1} \ll 1$, which is considerably smaller than the beaming angle of the jet associated with the GRB, the region of the blast wave actually accessible to a distant observer is then very small.
The emission along the line-of-sight axis to the source center suffers from the shortest geometric time delay, and hence originates at a larger radius (lower Lorentz factor) and is dimmer than slightly off-axis emission. The source therefore appears as a narrow limb-brightened ring \citep{limb}. The outer cutoff of the ring is set by the sharp decline of the relativistic beaming and due to the relativistic aberration the shock surface for a distance observer appears almost aligned with the line of sight at the edge of the ring. The small scale randomly generated magnetic field at the limb of the ring does not average out and some net polarization directed radially is possible. Since the source for a distant observer is symmetric the net polarization from such a source is expected to vanish unless the symmetry is broken, for instance due to polarization scintillation in the radio band or, as already mentioned, by gravitational microlensing \citep{micropol}. A late-time decline in the amplitude of intensity scintillations for radio afterglows has been detected and allowed us to derive direct constraints on the physical size of the emitting region \citep{radioscint,polradiofirst}. A rather detailed analysis of the expected polarization due to scintillation on radio observations of GRB afterglows was carried out by \citet{weibml}. At early times the source size is small compared to the characteristic angular scale of the scintillation and the expected polarization is low, however it grows monotonically with the source angular size and should saturate to the intrinsic polarization level emitted at the source (possibly that predicted by Eq.\,\ref{eq:synchroval}) in several weeks.

\begin{figure}
\begin{center}
\includegraphics[width=8cm]{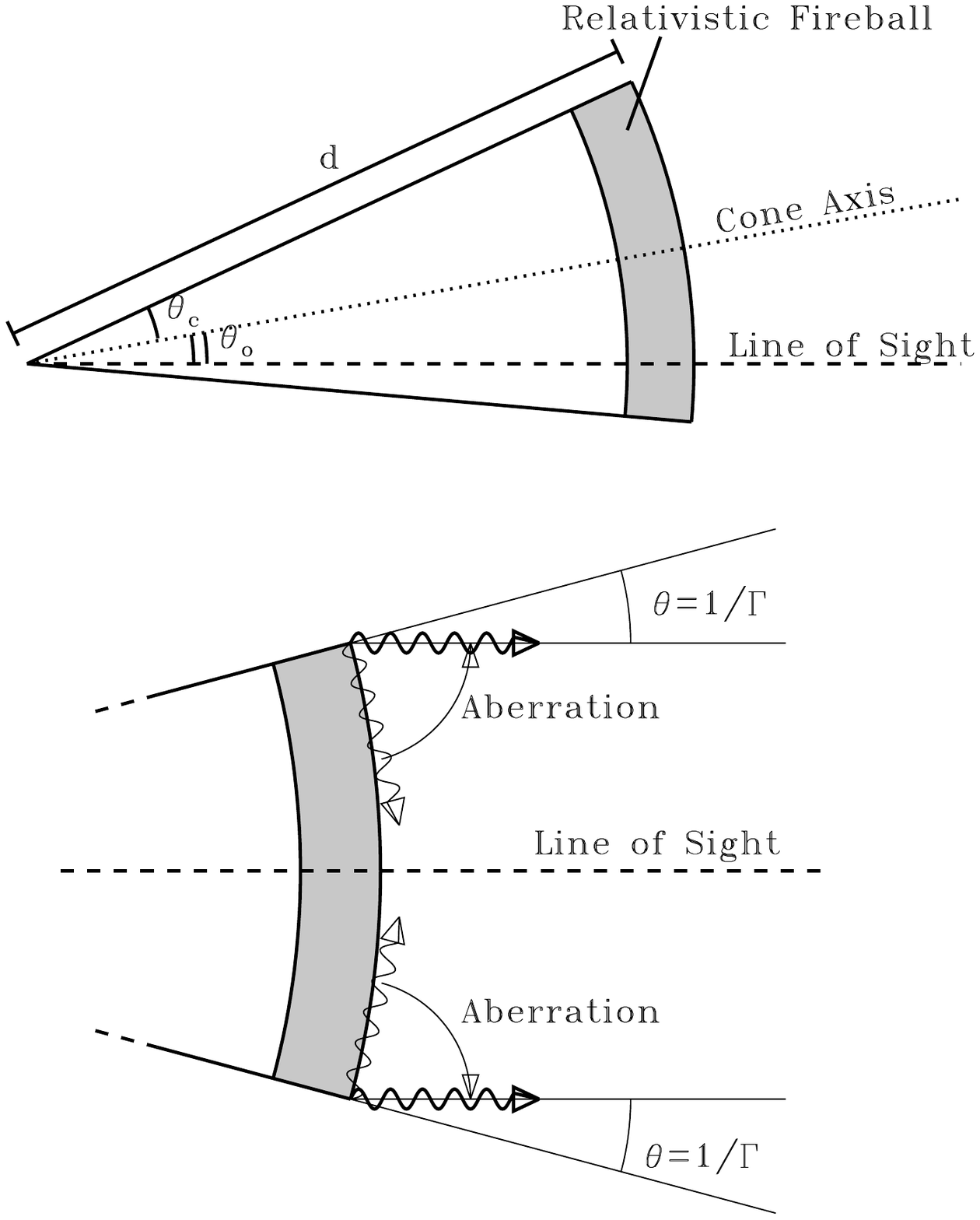}
\caption{Geometry of the beamed fireball. Note that photons emitted in the comoving frame at an angle $\pi/2$ from the velocity vector are those making an angle $\theta \sim 1/\Gamma$ with the line of sight in the observer frame. From \citet{ghisellini99}.}
\label{fig:aberration}
\end{center}
\end{figure}

A different, and in principle complementary, approach was developed almost simultaneously by \citet{ghisellini99} and \citet{sarippm}. The idea is based on the assumption that we are seeing a collimated fireball, i.e. a jet, slightly off-axis. Even in case the locally generated magnetic field at the shock is completely tangled, the anisotropy introduced by the interplay between the physically collimated emitting region and the aberration due to the ultra relativistic motion of the shock front can introduce some linear polarization. The magnetic field is assumed to be completely tangled if the shock is observed face-on, but with some degree of alignment if observed edge-on. This might happen for the magnetic configuration discussed in \citet{weibml} and \citet{patchybis} but also due to the effect of compression along one direction of the shocked region as proposed by \citet{compressed}. Photons emitted at right angle in the shock comoving frame can be polarized at a level, $P_0$, depending on the degree of order of the magnetic field in the plane perpendicular to the shock front. Since the emitting region is supposed to move with $\Gamma \gg 1$ these photons can then reach the observer (Fig.\,\ref{fig:aberration}). If $\theta_{\rm c}$ is the outflow opening angle and $\theta_0$ is the angle between the line of sight and the jet axis ($\theta_0 \le \theta_{\rm c}$), we can identify three regimes for the polarization depending on the time evolution of the outflow Lorentz factor. At early times $\Gamma$ is sufficiently high to allow the observation of a small area of the emitting region and the situation is perfectly symmetric, no or very small polarization should be observable. This is also true at late-time when the area accessible to the observer is large enough to include the whole outflow cone. At intermediate times, apart from the null probability case of line of sight perfectly aligned with the jet axis, $1/\Gamma$ becomes comparable to $\theta_{\rm c} - \theta_0$, and the observer begins to see the physical edge of the collimated outflow. The global symmetry is broken and some polarization is observed. This geometric model allows one to derive the polarization time-evolution since depending on the percentage of the outflow border visible for the observer the horizontal and vertical polarization component mix in a different way. Two polarization maxima are then expected (the first is due to the horizontal component, and the latter to the vertical one) with a period of null polarization in between. Between the two maxima a sharp rotation of the position angle by $90^\circ$ is predicted. The second maximum is always larger than the first and, according to \citet{ghisellini99}:
\begin{equation}
P_{\rm max} \simeq 0.19 \, P_0 \, \left( {\theta_o \over \theta_c} \right)^{2}.
\label{eq:approx}
\end{equation}
The above relation is true within a few percent if $1/ 20 \le (\theta_o / \theta_c) \le 1$ and $1^\circ \le \theta_c \le 15^\circ$. \citet{bjlindvar} explored the effect of a possible lateral expansion of the jet, which effectively translates into a change of the $\theta_o / \theta_c$ ratio. The authors argued that decreasing the ratio shifts the maxima toward later times and decreases their magnitude.

The specific polarization predictions depend on the detail of the deceleration of the fireball and on the values of $\theta_{\rm c}$ and $\theta_0$ (and $P_0$). However the general picture is independent of the model parameters and a strong link between the total flux from the afterglow and polarization can be singled out. The epoch corresponding to the position angle rotation should roughly coincide with the jet-break occurrence, i.e. when a distant observer realizes that the source is not spherical symmetric and records a deficit in emitting area that translates to a steeper decline of the light-curve. Therefore, the total and polarized flux time evolution should be closely linked to each other offering a powerful observational test for the model and the fireball parameters. 

The possible effect of an ordered field in the ambient medium for the observed polarization from GRB afterglows was investigated by \citet{granot03b}. The rationale was to possibly explain observations showing a constant or slowly variable polarization level during the afterglow evolution with almost constant position angle. We have already mentioned that for typical ISM the post-shock field would be too weak to produce the observed synchrotron emission. However, it could be stronger if the shock propagates into a magnetized wind of a progenitor star or into a pulsar-wind bubble (the magnetic energy fraction, $\epsilon_B$, would increase from about $10^{-10}$ to $10^{-4} - 10^{-1}$). An ordered magnetic field would affect the observed polarization depending on ratio of the ordered-to-random field. The total polarization from an afterglow with a locally generated random magnetic field, $B_{\rm rnd}$, and an ordered component, $B_{\rm ord}$, turns out to be:
\begin{eqnarray}\label{P_tot}
P &=& \left({\eta P_{\rm ord}\over 1+\eta}\right) \nonumber \\ 
    &  & \left [
1+\left({P_{\rm rnd}\over\eta P_{\rm ord}}\right)^2
-2\left({P_{\rm rnd}\over\eta P_{\rm ord}}\right)\cos
2\delta\right ]^{1/2}\,
\\ \label{theta}
\theta &=& {1\over 2}\arctan\left(
{\sin 2\delta\over\cos 2\delta-\eta P_{\rm ord}/P_{\rm rnd}}\right)\,
\end{eqnarray}
where $\eta\equiv I_{\rm ord}/I_{\rm rnd}\approx \langle B_{\rm ord}^2\rangle/\langle B_{\rm rnd}^2\rangle$ is the ratio of the observed intensities in the two components and $\delta$ is the angle between the ordered field and the jet axis. Assuming that $P_{\rm ord}$ is close to the maximum theoretical polarization (Eq.\,\ref{eq:synchroval}) typically we have $P_{\rm ord} >> P_{\rm rnd}$, and the low values of the observed polarization degrees (Tables\,\ref{tab:pol}, \ref{tab:polcnt}, \ref{tab:030329pol}, \ref{tab:091018pol}, and \ref{tab:121024pol}) imply that $\eta << 1$. The time evolution of the polarization essentially depends on the time evolution of $\eta$, which in turn depends on the ambient medium, making possible a large variety of different behaviors. In general, until $\eta P_{\rm ord} > P_{\rm rnd}$, the changes in the position angle would be moderate, whereas the variation in $P$ could be significant.

In addition, the magnetic field in the GRB ejecta is potentially much more ordered than in the shocked ambient medium behind the afterglow shock,
reflecting the likely presence of a dynamically important, predominantly transverse, large-scale field advected from the source. This could generate a large polarization value during the very early afterglow if the emission is dominated by the reverse-shock \citep{revpir,revsh}.
 
A potentially important diagnostic of the existence of ordered magnetic fields is provided by the observation of circular polarimetry \citep{matiocirc}. Circular polarization could be intrinsic, i.e. due to the synchrotron emission of the afterglow or can be generated by plasma effects as Faraday conversion. The Faraday conversion can convert some of the linear polarization from a source to circular polarization and is effective for synchrotron sources close to the self-absorption frequency typically for the afterglows in the radio bands. The detailed analysis carried out by \citet{matiocirc} showed that if the magnetic fields are tangled the circular polarization vanishes, while if the ordered component of the magnetic field is at least comparable to the tangled one, circular polarization at about 1\% in the radio bands, and 0.01\% in the optical is expected. During the early reverse shock it is also possible to have circular polarization one order of magnitude larger even if the ordered magnetic field component is very weak. A deeper analysis of the plasma effects on the observed polarization was carried out by \citet{sagivplasma}, deriving results qualitatively analogous to \citet{matiocirc}, although based also on a more accurate treatment of the effects of synchrotron losses that give a higher degree of circular polarization close to the frequencies where Faraday conversion from linear to circular polarization is more effective. A detailed study of the circular to linear polarization ratio in typical GRB afterglow configurations was also recently carried out by \citet{navacirc}. Their main results show that it is possible to assume ``ad-hoc" configurations allowing a large local circular polarization. However, once transformations from the local to the observer frame and integration across the whole visible region are performed, the circular to linear polarization ratio always vanishes in any realistic optical thin synchrotron emission afterglow model. 

Observations in the radio band can be very effective not only for accessing the early afterglow on a more relaxed time-scale, but also by means of relatively later-time observations able to provide information about the electron-proton coupling in the relativistic collisionless shocks supposed to originate the GRB afterglows \citep{tomaradiopol}. The fraction of electrons that are coupled to protons and accelerated, $f$, is usually hidden in the fraction of the total energy that goes in accelerating the electrons during synchrotron emission, $\epsilon_{\rm e}$. This has important possible consequences in the evaluation of the energetic of the events that would be larger by a factor $f^{-1}$ compared to the case with perfect coupling, $f=1$. In case the coupling is ineffective, $f < 1$, there could be thermal electrons available and the effect of Faraday rotation on these thermal electrons \citep{tomaradiopol} may suppress the linear polarization of the afterglow at frequencies higher than the absorption frequency and below a characteristic frequency that depends on the electron-proton coupling fraction. This effect would therefore be measurable by means of radio polarization observations at different frequencies. This mechanism could however work only if the magnetic fields are globally ordered to some extent \citep{sagivplasma}, while if the field is random with short coherence length the Faraday depolarization does not occur \citep{matiocirc}.

\begin{figure*}
\begin{center}
\begin{tabular}{>{\centering\arraybackslash}m{6cm} >{\centering\arraybackslash}m{8cm}}
\includegraphics[width=5cm]{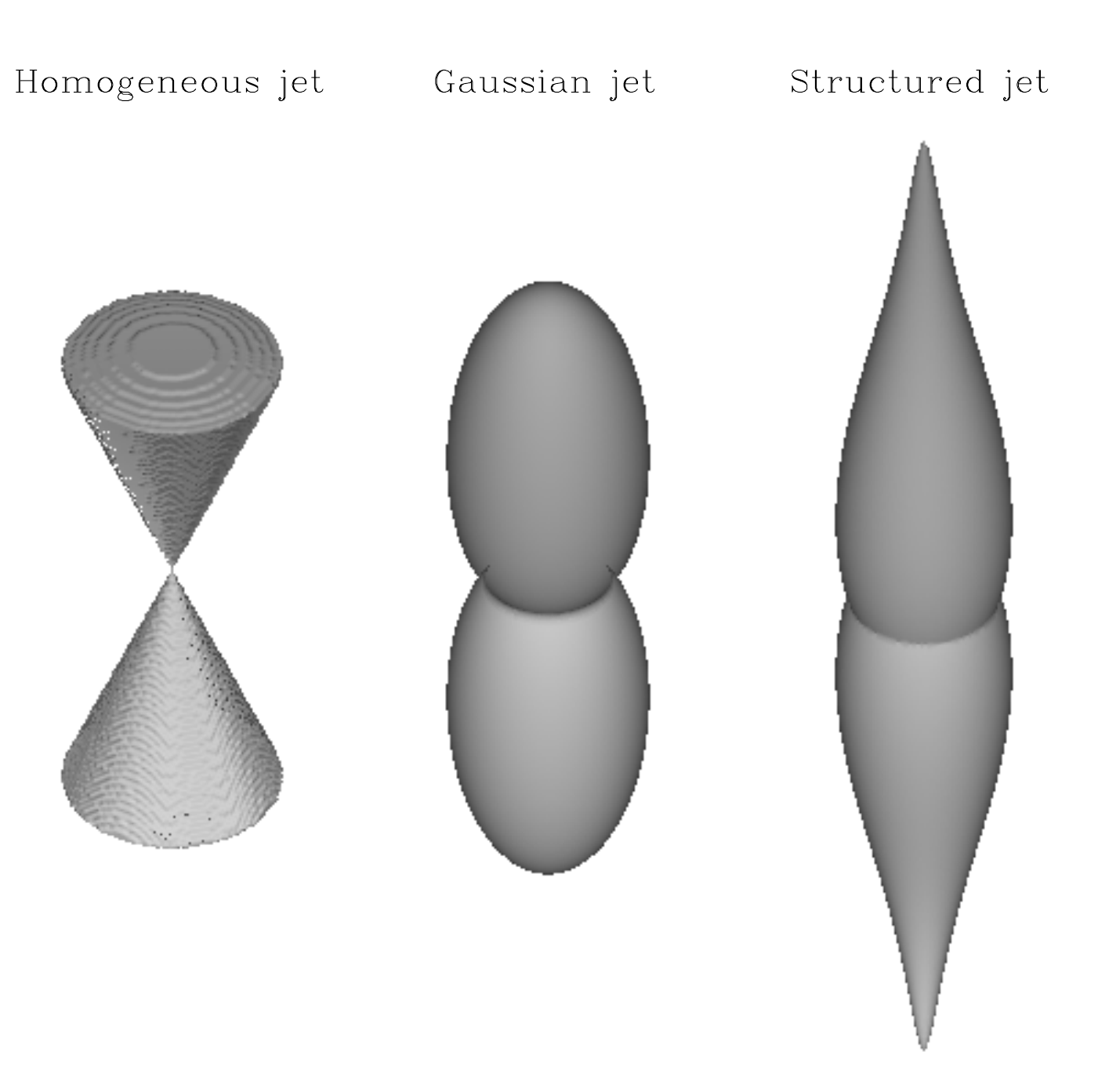} & \includegraphics[width=7.5cm]{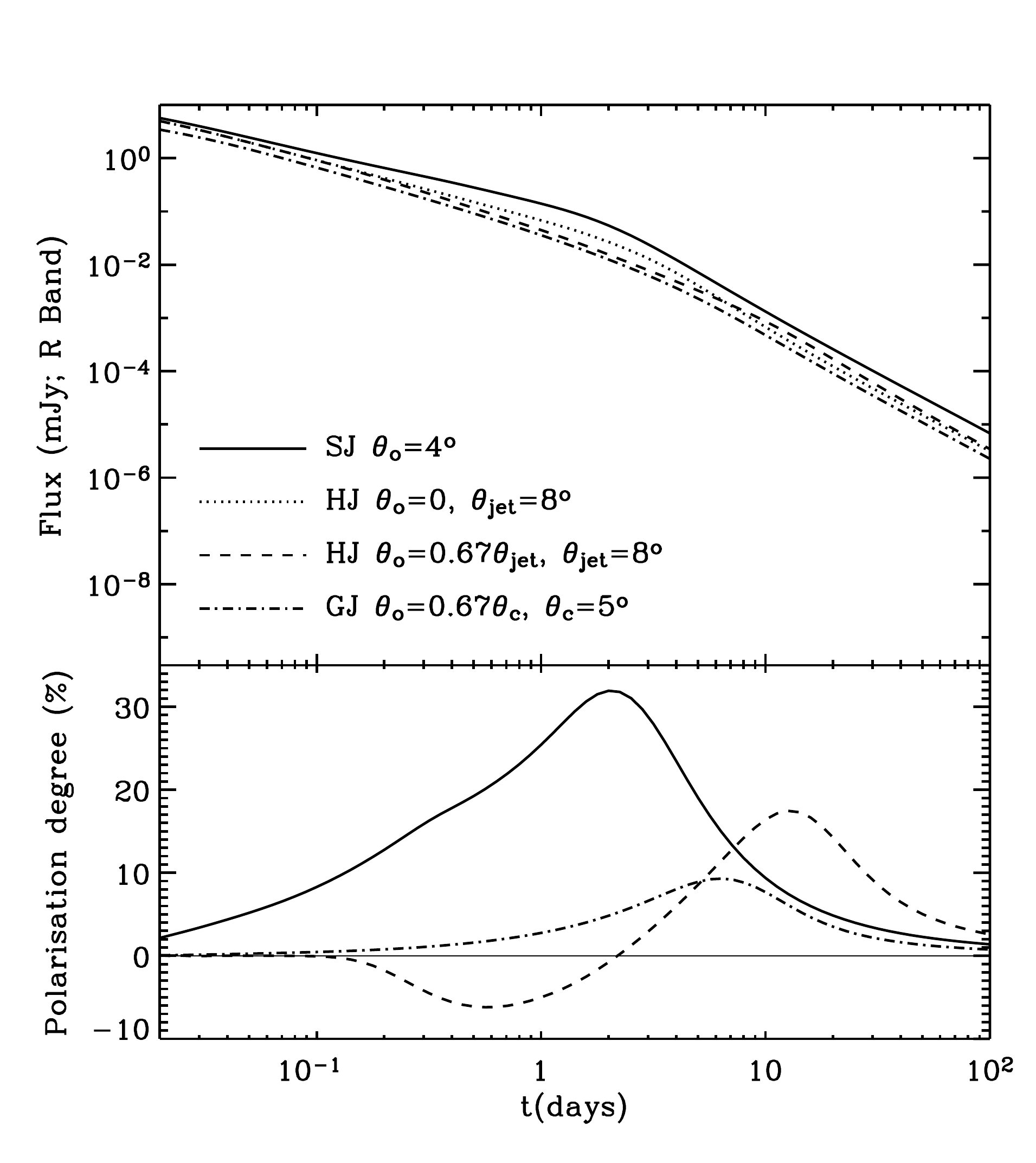} 
\end{tabular}
\caption{(\textit{left}) Three possible jet configurations. The figure shows the energy per unit solid angle of the jets logarithmically scaled. (\textit{right}) Light-curves (upper panel) and polarization curves (lower panel) comparison between a structured jet (SJ), a homogeneous jet (HJ) and a Gaussian jet (GJ) with given parameters. From \citet{rossipol}.}
\label{fig:jets}
\end{center}
\end{figure*}

The geometric models were originally developed assuming a homogeneous jet structure, i.e. at any given angle from the apex of the jet the luminosity emitted per unit solid angle along the jet axis and along the jet borders is the same. It is however of great interest to explore the possibility that the jet structure is more elaborated. Typical ideas can assume that the radiated power per unit solid angle is larger along the jet axis than along the wings, the so-called structured jets, or even that the jet luminosity follows a Gaussian distribution, the so-called Gaussian jets, with a core with almost constant luminosity that decreases exponentially outside of it (see Fig.\,\ref{fig:jets}). \citet{rossijet} and \citet{saljet} showed that the light-curves of the total flux from these configurations are very similar to each other. Things are, on the contrary, very different as far as linear polarization is concerned, suggesting the possibility that polarimetry could be a powerful diagnostic of the afterglow jet structure \citep{rossipol}. In general, for any off-axis observer, polarization is produced because different parts of the emitting jet surfaces do not contribute equally to the observed flux. In the homogeneous jet model this starts to occur when the emitting surface available to the observer includes the near border of the jet. In structured jet models the required asymmetry is intrinsic in the assumption that the emission depends on the angular distance from the jet axis. As it is shown in Fig.\,\ref{fig:jets}, based on the comprehensive analysis discussed in \citet{rossipol}, the predictions for different jet structures are markedly different. Structured jets show some (weak) polarization from the beginning, but the most important difference is for the $90^\circ$ rotation of the position angle predicted for homogeneous jets. Structured and Gaussian jets, since their emission is always dominated by the central core at the same angle with respect to the line of sight, do not show a position angle rotation with one only maximum for the polarization, typically close or after the jet-break time, depending on the specific jet parameters and structure.

In general, all the considerations discussed about the jet structure are based on single component jets. \citet{twojetswu} also considered the possibility of a two-component jet, with the inner component narrow and more energetic, and outer one wide and less energetic. The resulting light curves and polarization evolution depend strongly on the ratio of the intrinsic parameters of the two components, allowing a considerable freedom in modeling the observations.

The scenario depicted by the standard afterglow model can also be modified in case the axial symmetry is broken for instance if the energy per solid angle of the blast-wave display angular variations, the so-called ``patchy-shell'' model \citep{nakar&oren}. This idea was mainly developed to deal with the observations of GRB afterglows with fluctuations in their light-curves superposed to the general behavior predicted by the standard afterglow model. The variations in the degree and angle of polarization are here correlated to the light-curve variability.

Although observationally very demanding, the early afterglow has always attracted a considerable attention for its relevant diagnostic power. \citet{fanearly} analyzed the reverse shock emission powered by a magnetized outflow, possibly generated if the progenitor is magnetized and the field is dissipated. The reverse shock can produce a considerably bright optical flash depending on the magnetization parameter, $\sigma$ (the ratio of the electromagnetic energy to the particle energy) and the circumburst matter density profile. As a general rule, relatively low values for $\sigma \sim 0.05-0.1$ give the brightest optical flashes. In the so-called ICMART model \citep{icmart} $\sigma$ is indeed predicted to be after the magnetic dissipation close to unity based on energy equipartition arguments. As it was already pointed out \citep[e.g.,][]{granot03b}, the net linear polarization, $\Pi$, resulting from these bright optical flashes depend on the ratio between the ordered and random magnetic fields, $b$, and during the reverse shock \citet{fanearly} computed $\Pi \simeq 0.6\frac{b^2}{1+b^2}$. For low $\sigma$ the corresponding toroidal magnetic field is stronger than that generated at the shock and $b >> 1$, giving a very high local polarization. In case of an ultra-relativistic ejecta, due to beaming effects, only a small portion of the of the emitting region is visible. However, if the line of sight is even slightly off-axis the symmetric axis of the ordered magnetic field high polarization is still expected since the average of regions with different magnetic field directions is not effective. A very detailed study of the polarization in the early afterglow was also produced by \citet{lanearly}. Reverse- and forward-shock contributions are considered, and the hydrodynamics of the ejecta are computed in the case of thick and thin shells. The authors assumed that for the forward-shock the generated magnetic field is mainly random, while in the reverse-shock region the magnetic field can be large-scale ordered. This field is carried out from the central engine but later magnetic dissipations during the prompt GRB phase may reduce the magnetization degree to a lower level so that the ejecta is dominated by baryons and leptons in the afterglow phase but the large-scale structure of the magnetic field remains. The polarization evolutions of the early afterglows are mainly determined by the detailed magnetic field configurations and two field configurations were considered: toroidal and aligned with the jet axis. The magnetic field configurations can be associated to different progenitors, i.e. an aligned configuration would point to a magnetar, while a toroidal field is possibly indicating a black-hole. As a general conclusion, if the emission is dominated by the forward-shock, the polarization is expected to be very low in particular at early times. If the reverse-shock dominates, the polarization can reach $\sim 30-50$\%. It is always possible, however, to have configurations with a high level of symmetry not yielding detectable polarization. If the line of sight is outside the jet cone, a peak in the polarization degree is predicted with also an evolution of the position angle. An abrupt rotation of the position angle by $90^\circ$ is expected for some configurations around the reverse-shock shell crossing time. 

Finally, a complementary way to derive information about GRB emission ejecta magnetization would be the polarimetric study of X-ray flares \citep{chincflares} as discussed in \citet{fanflares,fanxray} and \citet{fanxrayrev}. In general, X-ray polarimetry of the afterglow would open exciting diagnostic possibilities \citep[see also][]{lanshallow}. The different temporal behavior of the optical and X-ray afterglows observed  in many events suggested the possibility that X-ray emission traces also a prolonged activity of the central engine while the optical afterglow is more related to the real forward-shock emission \citep{ghisxray}. This would make the polarization behavior of the optical and X-ray afterglows at least partially independent and with the capability to explore different emission regions of the GRB phenomenon. As discussed earlier, the early optical afterglow is expected to be weakly polarized unless a large scale ordered magnetic field is present. X-ray flares and plateau, as argued by \citet{fanearly}, might be driven by Poynting-flux dominated emission, so possibly showing a high level of linear polarization.

\subsection{Observations}
\label{sec:aftobs}

The first identification of a GRB afterglow was obtained for GRB\,970228 \citep{afterfirstx,afterfirsto}, although for GRB\,940217 a long-lasting high-energy emission possibly to be associated to the afterglow phase was already detected \citep{afterfirstg}. 

The first linear polarimetric observations (see Tables\,\ref{tab:pol}, \ref{tab:polcnt}, \ref{tab:030329pol}, \ref{tab:091018pol}, and \ref{tab:121024pol}) were carried out in the radio band with the VLA\footnote{http://www.vla.nrao.edu} about three weeks after GRB\,980329 \citep{polradiofirst}, yielding a rather shallow 21\% upper limit and about a week after the event for GRB\,980703 with 8\% upper limit \citep{FrGCN}. These limits are substantially lower than the theoretical synchrotron emission value (Eq.\,\ref{eq:synchroval}). However, the interpretation of the result is not direct since it depends on the location of the observing frequency compared to the synchrotron self-absorption frequency. At frequencies lower than the absorption frequency any intrinsic polarization is expected to be smeared out. The idea that the GRB afterglow emission is due to synchrotron emission was consistent with the broad-band spectral energy distribution observed for this event \citep{synchr}, but the debate was still open at that time.

Much more stringent was the $2.3$\% upper limit for linear polarization obtained by \citet{optupp} for the bright GRB\,990123 in the optical with the NOT\footnote{http://www.not.iac.es}. The observations were performed about 18 hours after the event. Such a low value was interpreted as possibly due to a jetted geometry with the GRB observed close to the beam axis. In a spherical geometry, again under the hypothesis that the emission was due to synchrotron radiation, such a low level requires highly tangled magnetic fields confined to the shock front \citep{optupp}. Milder upper limits for circular (see Table\,\ref{tab:cirpol}) and linear polarization were also obtained in the radio with the VLA \citep{kul990123,radiopol}.

\begin{figure}
\begin{center}
\includegraphics[width=8cm]{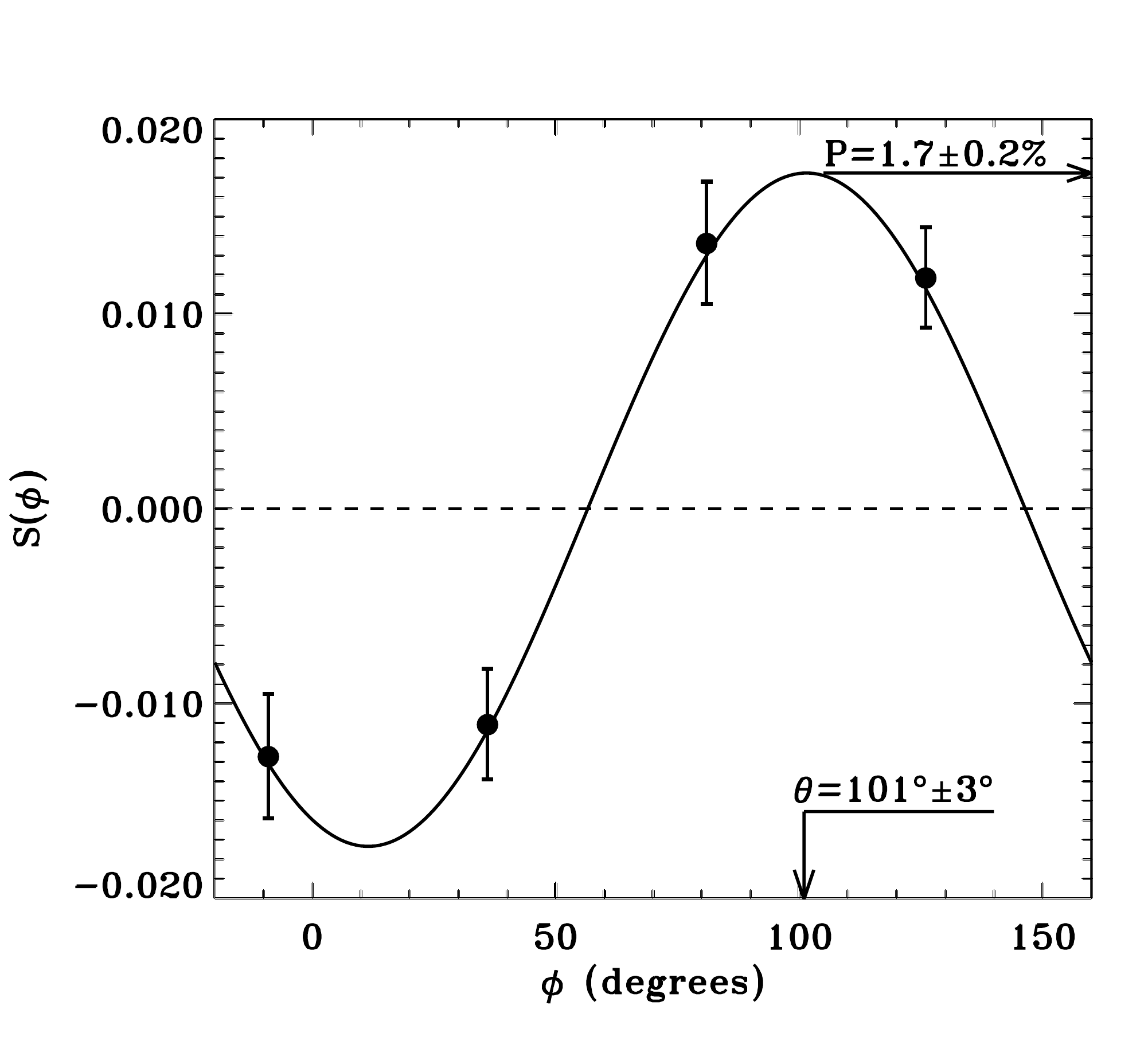}
\caption{Polarization measurement for GRB\,990510 obtained at the ESO-VLT. This is the first positive detection of polarization from a GRB afterglow. From \citet{afterfirstopt}. The function $S(\phi)$ gives the degree and position angle of the observed polarization \citep{disere}.}
\label{fig:firstpol}
\end{center}
\end{figure}

Polarimetry at a few percent level can be demanding for rapidly fading sources as GRB afterglows, and therefore it is not a surprise that the first positive detections in the optical band came a few months after the first unit of the VLT, with its collecting area and flexibility, became operational. GRB\,990510 was observed two times by two independent teams \citep{afterfirstopt,afterfirstoptbis} at 18-21\,hours after the burst with the ESO-VLT,\footnote{http://www.eso.org/public/teles-instr/vlt/} providing a small but highly significant polarization level at $P = 1.7 \pm 0.2$\% (Fig.\,\ref{fig:firstpol}). A later measurement one day after gave a result consistent with a non-variability of the observed polarization.

Polarization at this level is not common for extragalactic sources, however it is possible that it is due to polarization induced by dust grains interposed along the line of sight, which may be preferentially aligned due to the galactic magnetic fields. The effect is well known in our Galaxy \citep{mwpol} where it is observed that dust-induced polarization typically is in the range $P_{\rm max} \le 9.0E_{\rm B-V}$ and follows an empirically relation known as ``Serkowki law":
\begin{equation}
P(\lambda)/P_{\rm max} = e ^{K \ln^{2} (\lambda_{\rm max}/\lambda)},
\end{equation}
where $K \sim -1.15$ and $\lambda_{\rm max}$ in in the range 0.45-0.8\,$\mu m$ \citep{mwpol}. 

Large variations are anyway expected, and observed, for specific lines of sights. On the other hand, the effect due to dust in the Galaxy can be in principle easily checked and removed if a sufficiently large number of stars are observed in the same field of view. Since stars are typically intrinsically unpolarized (at least at this level of accuracy) any dust induced (as well as instrumental) polarization can be singled out. Dust-induced polarization in other galaxies has been studied only in a limited number of nearby cases \citep[e.g.,][]{m31pol} with results partly different compared to the Milky Way. For the case of GRB\,9905010, a sizable effect of dust induced polarization in the host galaxy could be ruled out since multi-band observations of the afterglow showed it could be effectively modeled as a power-law, according to the requirements of synchrotron shock model \citep{ssmod}, with essentially no rest-frame extinction \citep{afterfirstopt}.

Electron scattering can also lead to some polarization, as observed in supernovae and usually attributed to asymmetries in their photospheres \citep{snpol}. However, the degree of induced polarization is of the order of the electron optical depth that cannot be more than $\sim 10^{-6}$ a day after the event \citep{afterfirstoptbis}. We mention in passing that electron scattering as source of some polarized flux in GRB afterglows was also proposed by \citet{gneelectr} studying the polarization effects of radiation scattered in conical thin plasma envelopes. With different assumptions about the magnetic field geometry and the inclination of the cone axis with respect to the line of sight, it was possible to obtain the polarization at a few per cent level observed in GRB afterglows as due to Thomson scattering in the plasma of the conical jet. In this scenario synchrotron radiation does not contribute to the observed flux, and the observed polarization position angle should be essentially constant in time. 

GRB\,990510 was also the first GRB with an achromatic (at least in the optical band) steepening, a jet-break, of the afterglow light-curve clearly observed \citep{firstbreak,firstbreakbis,firstbreakter}. The observations with solid polarization detections were performed before the jet-break. Due to the large Lorentz factor of the outflow, $\Gamma$, only a fraction $\sim 1/\Gamma$ of the emitting region is accessible to the observer. Photons produced in regions at an angle $1/\Gamma$ with respect to the line of sight are emitted, in the comoving frame, at $\sim 90^\circ$ from the velocity vector. A comoving observer at this angle can then see a compressed emitting region \citep{ghisellini99} and a projected magnetic field structure with a preferred orientation. If the gradual steepening of the light curve is a jet-break, we would observe only regions at a viewing angle $1/\Gamma$ at variance with an axis-symmetric situation, and this asymmetry can be the cause of the observed linear polarization that therefore becomes the ``smoking gun" of synchrotron emission for GRB afterglows.

A few months later, for GRB\,990712, further evidence for the intrinsic origin of the observed afterglow polarization was obtained by \citet{rolvar}. Three epochs of linear polarimetric observations from about 10 to 35\,hr from the high energy event were obtained. The polarization was always in the 1-3\% range, but showed some variability with the minimum at the second epoch, while the position angle remained constant. The evidence for variability was only slightly better than $3\sigma$, yet no color variation of the afterglow was identified during the time period covered by the observations, making the possibility that the observed variation was due to dust in the host galaxy even less likely. The constancy of the position angle on such a long, compared to the afterglow evolution, monitoring is in disagreement with the hypothesis that the observed polarization could be due to a random mix of highly polarized emissions from independent magnetic domains \citep{patchy,patchybis}. While it is possible to have a low polarization assuming a large number of magnetic domains, the position angle should typically vary with the polarization during the afterglow evolution. And basically the same objection holds for polarization due to micro-lensing. The observations for GRB\,990712 were carried out during a phase of regular decay of the afterglow, i.e. no break was detected. The quality of the last observation of this dataset, which is consistent with both the first or the second observations due to the larger error bars, does not allow us to draw any further specific test based on the idea that the afterglow polarization is described by a geometric model  \citep{ghisellini99,sarippm} and these observations are generally consistent with that scenario, as also discussed in \citet{bjlindvar}.

Only upper limits were instead obtained for GRB\,991216, observed in the optical with the VLT \citep{covearlyrev} and in the radio with the VLA \citep{radiopol}. Being the most stringent limit at 2.7\%, it is possible that for this event the entire polarization evolution was characterized by lower values. It became indeed immediately clear that the best (largest) available facilities were required to derive effective polarimetry at 1\% level for GRB afterglows. Nevertheless, attempts were carried out with smaller size facilities or in the NIR, where the observing conditions are more difficult for polarimetry. GRB\,000301C was observed in the $K$ band with the Calar Alto 3.5m telescope\footnote{http://www.caha.es}, and only a mild upper limit at 30\% was obtained \citep{stenir}. GRB\,010222 was instead observed with the NOT about 23\,hours from the GRB in the $V$ band yielding a low significance detection at $1.36 \pm 0.64$\% \citep{bjeninj}. However, the low average polarization level typically detected in late-time GRB afterglows often resulted in upper limits with the largest facilities too, in particular when the observations could not be carried out before about one day from the high-energy event. GRB\,011211 was observed about 35\,hours after the GRB with the VLT and an upper limit at $3\sigma$ of 2.7\% was derived \citep{pol01}. These limits were poorly constraining yet generally consistent with the predictions of the geometric models \citep{ghisellini99,sarippm}.

The attempt to identify a time-evolution of the polarization degree and position angle generated a richer dataset for several events. GRB\,020405 was observed with the VLA \citep{radiopol}, the VLT \citep{mas020405,pol020405} and the Multiple Mirror Telescope \citep[MMT\footnote{http://www.mmto.org}, ][]{ber020405} between one and three days from the burst. The polarization level was observed at about 1.2-2\% for the VLT observations but at the MMT a much higher polarization at about 10\% was detected. Only mild upper limits were obtained at the radio frequencies. The position angle possibly showed a slow change ($\sim 10^\circ$) from the first to the last observations. GRB\,020405 is characterized by a large galaxy superposed to the afterglow position, but during the polarimetric campaign the afterglow was much brighter than the host galaxy and therefore able to only slightly affect the measurements \citep{pol020405}. During the observations the afterglow light-curve showed a regular and smooth decay \citep{mas020405}. The observations derived with the VLT before and after the MMT observations are substantially consistent with a constant afterglow polarization, possibly also with an important contribution of dust in the host galaxy. The rapid variation required to move from the $\sim 1$\% to $\sim 10$\% in a timescale of about one hour is essentially inconsistent with basically all the geometric models and also with the patchy-shell idea \citep{ghisellini99,sarippm,patchy}. In principle a micro-lensing phenomenon \citep{micropol} could be responsible for the polarization "flare", although the rapid time scale, the almost constant position angle and the lack of an analogous brightening in the total flux curve make even this interpretation unlikely. The high polarization observed by \citet{ber020405} only about one hour after the observation carried out by \citet{mas020405} is therefore still of difficult interpretation.

\begin{table*}
\caption{Linear polarization measurements carried out for several GRB afterglows. Partial collections of data are also reported in \citet{bjpol} and \citet{covearlyrev}. Data are corrected for Galactic induced polarization when available in the original references. More data are reported in Table\,\ref{tab:polcnt} and data for the large datasets of GRB\,030329, GRB\,091018 and GRB\,121024A are reported in Tables\,\ref{tab:030329pol}, \ref{tab:091018pol} and \ref{tab:121024pol}.}
\begin{footnotesize}
\begin{center}
\begin{tabular}{lcccccll}
\hline
\hline
Event & $t-t_0$ & P$_{\rm lin}$ & $\Theta$ & $\nu$ & Instrument & z & Ref \\
          & (hour)  & (\%)       & (deg)       & (Hz)  \\
\hline
GRB\,980329 & 500 & $< 21$ ($2\sigma$) &  & $8.3\times10^9$ & VLA & 3.5 & \citep{polradiofirst} \\
GRB\,980703 & 100     & $< 8$ ($3\sigma$) &  & $4.86\times10^9$ & VLA & 0.97 & \citep{FrGCN} \\
		      & 100      & $< 8$ ($3\sigma$) &  & $8.46\times10^9$ & VLA &  & \citep{FrGCN} \\
GRB\,990123 & 18.25  &  $< 2.3$ ($2\sigma$) &  & $4.55\times10^{14}$ & NOT & 1.6 & \citep{optupp} \\
		      & 30.0  & $< 23$ ($3\sigma$) & & $8.46\times10^9$ & VLA & & \citep{kul990123} \\
		      &		&						&	&			&	&	& \citep{radiopol} \\
GRB\,990510 & 18.5 & $1.7 \pm 0.2$ & $101 \pm 3$ & $4.55\times10^{14}$ & VLT & 1.62 & \citep{afterfirstopt} \\
                       & 20.6 & $1.6 \pm 0.2$ & $96 \pm 4$ & $4.55\times10^{14}$ & VLT &  & \citep{afterfirstoptbis} \\
                       & 43.4 & $2.2^{+1.1}_{-0.9}$ & $112^{+17}_{-15}$ & $4.55\times10^{14}$ & VLT &  & \citep{afterfirstoptbis} \\       
GRB\,990712 &  10.56 & $2.9 \pm 0.4$ & $121.1 \pm 3.5$ & $4.55\times10^{14}$ & VLT & 0.43 & \citep{rolvar} \\  
                       &  16.8 & $1.2 \pm 0.4$ & $116.2 \pm 10.1$ & $4.55\times10^{14}$ & VLT &  & \citep{rolvar} \\ 
                       &  34.8 & $2.2 \pm 0.7$ & $139.1 \pm 10.4$ & $4.55\times10^{14}$ & VLT &  & \citep{rolvar} \\  
GRB\,991216 &  35.0 & $< 2.7$ ($2\sigma$)  &  & $5.45\times10^{14}$ & VLT & 1.02 & \citep{covearlyrev} \\   
		      & 35.8 &  $< 11$ ($3\sigma$)  & & $8.46\times10^9$ & VLA & & \citep{radiopol} \\
  	              &  60.0 & $< 5$ ($2\sigma$)  &  & $5.45\times10^{14}$ & VLT &  & \citep{covearlyrev} \\     
		      & 64.3 &  $< 9$ ($3\sigma$)  & & $8.46\times10^9$ & VLA & & \citep{radiopol} \\  
GRB\,000301C & 43 &  $< 30$  & & $4.55\times10^{14}$ & VLT & 2.03 & \citep{stenir} \\
GRB\,010222 & 22.65 &  $1.36 \pm 0.64$  & & $5.45\times10^{14}$ & NOT & 1.48 & \citep{bjeninj} \\
GRB\,011211 & 35 &  $< 2.7$ ($3\sigma$)  & & $4.55\times10^{14}$ & VLT & 2.14 & \citep{pol01} \\
GRB\,020405 & 28.6 & $< 11$ ($3\sigma$)  & & $8.46\times10^9$ & VLA & 0.69 & \citep{radiopol} \\
		      & 29.5 & $1.50 \pm 0.40$ & $172 \pm 8$ & $4.55\times10^{14}$ & VLT &  & \citep{mas020405} \\	
		      & 31.7 & $9.89 \pm 1.3$ & $179.9 \pm 3.8$ & $5.45\times10^{14}$ & MMT &  & \citep{ber020405} \\
		      & 52.0 & $1.96 \pm 0.33$ & $154 \pm 5$ & $5.45\times10^{14}$ & VLT &  & \citep{pol020405} \\
		      & 76.2 & $1.47 \pm 0.43$ & $168 \pm 9$ & $5.45\times10^{14}$ & VLT &  & \citep{pol020405} \\
GRB\,020813 & 4.7-7.9 & $1.8-2.4$ & $148-162$ & $3.26-9.37\times10^{14}$ & Keck & 1.25 & \citep{bar020813} \\ 
		      & 21.55 & $1.07\pm0.22$ & $154.3\pm5.9$ & $5.45\times10^{14}$ & VLT &  & \citep{goros020813} \\
		      & 22.5 & $1.42\pm0.25$ & $137.0\pm4.4$ & $5.45\times10^{14}$ & VLT &  & \citep{goros020813} \\
		      & 23.41 & $1.11\pm0.22$ & $150.5\pm5.5$ & $5.45\times10^{14}$ & VLT &  & \citep{goros020813} \\
		      & 24.39 & $1.05\pm0.23$ & $146.4\pm6.2$ & $5.45\times10^{14}$ & VLT &  & \citep{goros020813} \\
		      & 26.80 & $1.43\pm0.44$ & $155.8\pm8.5$ & $5.45\times10^{14}$ & VLT &  & \citep{goros020813} \\
		      & 27.34 & $1.07\pm0.53$ & $163.0\pm14.6$ & $5.45\times10^{14}$ & VLT &  & \citep{goros020813} \\
		      & 27.78 & $1.37\pm0.49$ & $142.1\pm8.9$ & $5.45\times10^{14}$ & VLT &  & \citep{goros020813} \\
		      & 47.51 & $1.26\pm0.34$ & $164.7\pm7.4$ & $5.45\times10^{14}$ & VLT &  & \citep{goros020813} \\	
		      & 97.29 & $0.58\pm1.08$ & $13.7\pm24.4$ & $5.45\times10^{14}$ & VLT &  & \citep{goros020813} \\	
\hline
\end{tabular}
\end{center}
\end{footnotesize}
\label{tab:pol}
\end{table*}

\begin{table*}
\caption{Linear polarization measurements carried out for several GRB afterglows. Continued from Table\,\ref{tab:pol}}
\begin{footnotesize}
\begin{center}
\begin{tabular}{lcccccll}
\hline
\hline
Event & $t-t_0$ & P$_{\rm lin}$ & $\Theta$ & $\nu$ & Instrument & z & Ref \\
          & (hour)  & (\%)       & (deg)       & (Hz)  \\
\hline
GRB\,021004 & 8.88   & $1.88\pm0.46$ & $189\pm7$ & $4.55\times10^{14}$ & NOT & 2.33 & \citep{rol021004} \\
		      & 9.12   & $2.24\pm0.51$ & $175\pm6$ & $4.55\times10^{14}$ & NOT &  & \citep{rol021004} \\
		      & 9.60   & $< 0.60$ & & $4.55\times10^{14}$ & NOT &  & \citep{rol021004} \\
		      & 10.76   & $< 5.0$ ($2\sigma$) &  & $2.43\times10^{14}$ & TNG &  & \citep{laz021004} \\	
		      & 14.62   & $0.51\pm0.10$ & $126\pm5$ & $5.45\times10^{14}$ & VLT &  & \citep{laz021004} \\	
		      & 16.08   & $0.71\pm0.13$ & $140\pm5$ & $5.45\times10^{14}$ & VLT &  & \citep{rol021004} \\    
		      & 18.83  & $0.8-1.7$ & $100-147$ & $3.49-8.57\times10^{14}$ & VLT &  & \citep{laz021004} \\
		      &		&		&				&				&	&	& \citep{wang021004} \\ 
		      & 90.7  & $0.43\pm0.20$ & $45\pm12$ & $5.45\times10^{14}$ & VLT &  & \citep{laz021004} \\
GRB\,030226 & 25.39 & $< 1.1$ ($2\sigma$) & & $4.55\times10^{14}$ & VLT & 1.99 & \citep{klo030226} \\
GRB\,030328 & 18.5 & $2.4\pm0.6$ & $170\pm7$ & $5.45\times10^{14}$ & VLT & 1.52 & \citep{mai030328} \\
GRB\,060418 & 0.057 & $< 8$ ($2\sigma$) & & $5.08\times10^{14}$ & LT & 1.49 & \citep{mund060418} \\
GRB\,071010A & 21.51 & $< 1.3$ ($3\sigma$) & & $4.55\times10^{14}$ & VLT & 0.99 & \citep{cov071010} \\
GRB\,080310 & 24.21 & $< 2.5$ ($2\sigma$) & & $5.45\times10^{14}$ & VLT & 2.43 &  \citep{littlejohns080310} \\
			& 47.08 & $< 2.5$ ($2\sigma$) & & $5.45\times10^{14}$ & VLT &   &  \citep{littlejohns080310} \\
			& 70.48 & $< 2.6$ ($2\sigma$) & & $5.45\times10^{14}$ & VLT &   &  \citep{littlejohns080310} \\			
GRB\,080928 & 15.2 & $2.5\pm0.5$ & $27\pm3$ & $4.25-6.00\times10^{14}$ & VLT & 1.69 & (this paper) \\
GRB\,090102 & 0.045 & $10.1\pm1.3$ &  & $5.08\times10^{14}$ & LT & 1.55 & \citep{ste090102} \\
GRB\,091208B & 0.10 & $10.4\pm2.5$ & $92 \pm 6$ & $4.55\times10^{14}$ & Kanata & 1.06 & \citep{ue091208} \\
GRB\,100906A & $\sim 0.5$ & $<10$ ($60$\%) & & $\sim 5\times10^{14}$ & MASTER & 1.73 & \citep{gorb100906} \\
GRB\,110205A & 0.0675 & $< 16$ ($3\sigma$) &    & $5.08\times10^{14}$ & LT & 2.21 & \citep{cucc2011} \\
			& 0.93 & $< 6.2$ ($2\sigma$) &    & $5.08\times10^{14}$ & LT &  & \citep{cucc2011} \\
			& 3.53 & $1.4$ & & $4.55\times10^{14}$ & CAHA & & \citep{goros110205} \\
GRB\,120308A & 0.0292 & $28 \pm 4$ & $34 \pm 4$ & $5.08\times10^{14}$ & LT & & \citep{mundell13} \\
			& 0.0594 & $23 \pm 4$ & $44 \pm 6$ & $5.08\times10^{14}$ & LT & & \citep{mundell13} \\
			& 0.0892 & $17^{+5}_{-4}$ & $51 \pm 9$ & $5.08\times10^{14}$ & LT & & \citep{mundell13} \\
			& 0.1189 & $16^{+7}_{-4}$ & $40 \pm 10$ & $5.08\times10^{14}$ & LT & & \citep{mundell13} \\
			& 0.1792 & $16^{+5}_{-4}$ & $55 \pm 9$ & $5.08\times10^{14}$ & LT & & \citep{mundell13} \\
GRB\,121011A & $\sim 0.5$ & $<15$ ($60$\%) & & $\sim 5\times10^{14}$ & MASTER & & \citep{prumaster} \\
GRB\,130427 	& 36 &  $< 3.9$ ($3\sigma$) & & $4.8\times10^{9}$ & EVN & 0.34 & \citep{horst130427} \\
			& 60 &  $< 7.5$ ($3\sigma$) & & $4.8\times10^{9}$ & EVN &  &   \citep{horst130427} \\
			& 110 &  $< 21$ ($3\sigma$) & & $4.8\times10^{9}$ & EVN &  &   \citep{horst130427} \\
GRB\,131030A & 0.9 & $2.1 \pm 1.6$ & $27 \pm 22$ & $4.55\times10^{14}$ & Skinakas 1.3m & 1.29 & \citep{king131030} \\
GRB\,140430A & 0.051 & $< 22$ ($3\sigma$) & & $5.08\times10^{14}$ & LT & 1.60 & \citep{kopac15} \\
GRB\,150301B & 0.023 & $> 7.6$ &  & $5\times10^{14}$ & MASTER & 1.52 & \citep{gorbovskoy16} \\
\hline
\end{tabular}
\end{center}
\end{footnotesize}
\label{tab:polcnt}
\end{table*}

The reverse-shock emission in the radio band, as a potential probe of emission from the GRB ejecta, has attracted a considerable attention also because in the radio band the time evolution is slower than in the optical. Observations of radio flares are indeed often interpreted as due to the reverse-shock, and rather mild limits for linear and circular polarimetry in this scenario for three events, GRB\,990123, GRB\,991216, GRB\,020405, were presented and discussed by \citet{radiopol}. 

One of the most interesting datasets was instead obtained for GRB\,020813 \citep{goros020813,laz020813}. For this event it was possible to secure observations before and after the jet-break time with the ESO-VLT. Spectro-polarimetric observations were also obtained \citep{bar020813} with the Keck telescope\footnote{http://www.keckobservatory.org}. The linear polarization was slightly higher than 2\% at about 5-8\,hours after the GRB, and later was detected at $\sim 1$\% level with an almost constant (or weakly changing) position angle. In particular, no large rotation of the position angle was observed before and after the jet-break time (0.4-0.9\,days). GRB\,020813 was also characterized by a very smooth light-curve \citep{goros020813}, ensuring that inhomogeneities in the fireball or in the surrounding circumstellar medium are not important and therefore unable to significantly affect the polarization measurements. In \citet{laz020813} several possibilities were discussed, including scenarios with magnetized jets. An important result was that models based on homogeneous jets, implying a 90$^\circ$ rotation, are ruled out by the data. On the contrary, models described by structured jets, predicting a polarization peak close to the jet-break time \citep{rossijet,laz020813,rossipol}, are more consistent with the data. However, as suggested by the relatively high polarization value found at early times, models assuming a magnetized jet, i.e. a jet with a non-negligible toroidal component, are also able to satisfactorily fit the data both for homogeneous and structured jets. \citet{twojetswu} showed that GRB\,020813 polarimetric and photometric data could also be successfully modeled by a two-component jet with the line of sight within the wide component to ensure the constancy of the polarization position angle. Attempts to model the polarization evolution of GRB\,080203 were also proposed by \citet{dadopol}, in the context of their ``cannonball" model. Since the intrinsic afterglow emission from the fireballs, plasmoids ejected by the central engine of the GRB at very high Lorentz factor \citep[see][and references therein]{dadopol} should be unpolarized, the observed polarization is attributed to the effect of dust along the line of sight in the GRB host galaxies. The lack of sizable reddening in the total multicolor light-curve modeling is however difficult to reconcile with this scenario.

Multiple observations covering the afterglow evolution from about 0.3 to 3\,days after the high-energy event were obtained also for GRB\,021004. In this case the light-curve was characterized by several re-brightenings making the modeling more complex \citep{bjo021004} and not allowing in particular to single out unambiguously the jet-break time. Polarimetry was carried out with the NOT \citep{rol021004} at about 9\,hours after the GRB and with the ESO-VLT a few hours later. The polarization degree was approximately constant at $1-1.5$\% but the position angle changed by $\sim 45^\circ$ between the NOT and VLT observations. More polarimetric observations were presented by \citet{laz021004} obtained with the TNG\footnote{http://www.tng.iac.es} in the $J$ band $\sim 10$\,hours after the GRB and with the ESO-VLT in the optical between $\sim 14$ and $\sim 90$\,hours after the high-energy event. The TNG observation gave a 5\% upper limit while with the VLT the measured polarization decreased from $\sim 1.3$\% down to $\sim 0.7$\%, showing also a rotation of the position angle consistent with a gradual $90^\circ$ rotation with respect to the earliest measurements. Spectro-polarimetry, obtained with the ESO-VLT about 18\,hours after the GRB, showing a polarization $\sim 1.8$\% and position angle $\sim 118^\circ$, were presented and discussed by \citet{wang021004} and \citet{laz021004}. 

The typically low observed polarization and the relatively high redshift of the GRBs imply that any possible contribution due to dust in the host galaxy must be carefully considered. The possibility that the observed linear polarization from GRB afterglows could be just an artifact due to dust in the host galaxies was already discussed for other events and typically ruled out by the low level of observed reddening in multi-colour photometry. The contributions to the observed polarization due to dust in the Milky Way and in the host galaxy can be modeled, assuming a dichroic medium inducing a polarization $p_{\rm dust} \equiv q^2+u^2$ to background unpolarized sources, by a Mueller matrix \citep{polbook,disere} of the form:
\begin{equation}
M = e^{-\tau} 
\left( \begin{array}{cccc}
  1 & q & u & 0 \\
  q & {q^2+Au^2}\over{p_{\rm{dust}}^2} & {qu(1-A)}\over{p_{\rm{dust}}^2}  & 0  \\
  u & {qu(1-A)}\over{p_{\rm{dust}}^2}  & {u^2+Aq^2}\over{p_{\rm{dust}}^2} & 0  \\
  0 & 0        & 0       & A
\end{array} \right)
\label{eq:mue}
\end{equation}
where $A \equiv \sqrt{1-p_{\rm dust}^2}$, and $e^{-\tau}$ is the opacity of the medium to non-polarized radiation. For a deeper discussion about the range of validity of Eq.\,\ref{eq:mue}, see \citet{mueism}. 

Simulations carried out for the case of GRB\,021004 \citep{laz021004} showed that for low polarization levels a change of the position angle by a large amount due to the superposition of a varying, in intensity and polarization, source (the afterglow) and the effect of dust in the host galaxy is possible. However, a detailed fit based on the geometric model \citep{ghisellini99,sarippm} was difficult to achieve. The irregularities of the light-curve could possibly be due to inhomogeneities in the external medium or in the fireball itself. In both cases the break of the symmetry of the system can generate some polarization superposed to the general trend, making a reliable modeling dependent on too many free parameters considering the limited observational data.
As a matter of fact, \citet{bjo021004} showed that with the addition of a few episodes of energy injections the light-curve and the polarization evolution can satisfactorily be modeled assuming an homogenous jet structure. The effect of energy injection on polarization is due to the temporary increase of the fireball Lorentz factor $\Gamma$. Increasing $\Gamma$ the flux also increases, the relativistic aberration becomes more important and the emitting surface area decreases with a net result of a smaller polarization degree compared to an unperturbed situation. Rather interestingly, adopting a blast wave energy distribution lacking of axial symmetry allows one to obtain correlated light-curve and polarization degree and position angle variations, also able to satisfactorily model the GRB\,021004 data \citep{nakar&oren}.

For two more events only a few polarimetric measurements could be carried out. A stringent upper limit at 1.1\% was obtained for GRB\,030226 about one day after the GRB \citep{klo030226}. For GRB\,030328 polarization at $P \sim 2.4$\% was instead observed about 18\,hours after the GRB \citep{mai030328}, probably intrinsic to the event due to  the low local extinction as inferred by multi-color photometry and spectroscopy of the afterglow.

A fundamental breakthrough in the observational activities of GRB afterglows occurred with GRB\,030329 \citep{gre030329,mag030329,klo030329,radio030329,radio030329bis}. GRB\,030329 was discovered by the HETE\,II satellite\footnote{http://space.mit.edu/HETE/} and was one of the few cases of low redshift GRBs ($z\sim0.17$). Being in addition a regular cosmological GRB, i.e. not part of the category of low-luminosity low-redshift events \citep{pesll}, it showed an optical brightness sufficiently high to allow about one month of uninterrupted polarimetric observations in the optical (Table\,\ref{tab:030329pol}) with the ESO-VLT, the CAHA, the NOT, the IAG-USP\footnote{http://www.iag.usp.br/astronomia/} and much longer in the radio with the VLBA\footnote{https://public.nrao.edu/telescopes/vlba}. The afterglow polarization showed a strong variability in polarization degree and position angle. The polarization was typically in the 0.3-2.5\% range, and spectropolarimetry or multi-band observations showed that the position angle was constant in the optical and NIR bands. The light-curve of GRB\,030329 was characterized by numerous bumps and wiggles, and after about 10\,days a supernova component also affected the observations. The modeling of these polarization data is beyond the capabilities of any scenario discussed so far, lacking for instance of any clear correlation between polarization and light-curve behavior. Possibly, the observed emission and polarization is therefore due to the superposition of different phenomena that make a proper modeling difficult to achieve \citep{gre030329}. The vanishing radio polarization at late-times might be due to much less ordered than expected magnetic fields and/or Faraday rotation depolarizing the emission at the radio bands \citep{radio030329bis}.

\begin{table*}
\caption{Linear polarization measurements carried out for the afterglow of GRB\,030329.}
\begin{footnotesize}
\begin{center}
\begin{tabular}{lcccccll}
\hline
\hline
Event & $t-t_0$ & P$_{\rm lin}$ & $\Theta$ & $\nu$ & Instrument & z & Ref \\
          & (hour)  & (\%)       & (deg)       & (Hz)  \\
\hline
GRB\,030329 & 12.77 & $0.92 \pm 0.10$ & $86.13 \pm 2.43$ & $4.55\times10^{14}$ & VLT & 0.17 & \citep{gre030329} \\ 
 & 13.18 & $0.86 \pm 0.09$ & $86.74 \pm 2.40$ & $4.55\times10^{14}$ & VLT &  & \citep{gre030329} \\ 
 & 13.61 & $0.87 \pm 0.09$ & $88.60 \pm 2.64$ & $4.55\times10^{14}$ & VLT &  & \citep{gre030329} \\ 
 & 14.04 & $0.80 \pm 0.09$ & $91.12 \pm 2.88$ & $3.53-4.55\times10^{14}$ & VLT &  & \citep{gre030329} \\ 
 & 16.61 & $0.66 \pm 0.07$ & $78.52 \pm 2.94$ & $4.55\times10^{14}$ & VLT &  & \citep{gre030329} \\ 
 & 17.11 & $0.66 \pm 0.07$ & $76.69 \pm 2.89$ & $4.55\times10^{14}$ & VLT &  & \citep{gre030329} \\ 
 & 17.62 & $0.56 \pm 0.05$ & $74.37 \pm 3.11$ & $4.55\times10^{14}$ & VLT &  & \citep{gre030329} \\  
 & 35.19 & $2.10 \pm 1.20$ & $54.1 \pm 10.4$ & $1.39\times10^{14}$ & CAHA &  & \citep{klo030329} \\
 & 36.49 & $1.97 \pm 0.48$ & $83.2$ & $4.55\times10^{14}$ & IAG-USP &  & \citep{mag030329} \\ 
 & 36.72 & $1.10 \pm 0.40$ & $70 \pm 11$ & $4.55\times10^{14}$ & CAHA &  & \citep{gre030329} \\ 
  &		&			&				&				&		&	&	\citep{klo030329} \\ 
 & 37.20 & $1.37 \pm 0.11$ & $61.65 \pm 2.38$ & $4.55\times10^{14}$ & VLT &  & \citep{gre030329} \\ 
 & 37.92 & $1.50 \pm 0.12$ & $62.29 \pm 2.44$ & $4.55\times10^{14}$ & VLT &  & \citep{gre030329} \\ 
 & 40.08 & $1.07 \pm 0.09$ & $59.41 \pm 2.51$ & $3.53-4.55\times10^{14}$ & VLT &  & \citep{gre030329} \\ 
 & 40.80 & $1.09 \pm 0.08$ & $66.07 \pm 2.45$ & $4.55\times10^{14}$ & VLT &  & \citep{gre030329} \\
 &		&			&				&				&		&	&	\citep{klo030329} \\ 
 & 41.28 & $1.02 \pm 0.08$ & $67.05 \pm 2.60$ & $4.55\times10^{14}$ & VLT &  & \citep{gre030329} \\ 
 & 41.76 & $1.13 \pm 0.08$ & $70.56 \pm 2.51$ & $4.55\times10^{14}$ & VLT &  & \citep{gre030329} \\ 
 & 64.32 & $0.52 \pm 0.06$ & $30.76 \pm 5.04$ & $3.53-4.55\times10^{14}$ & VLT &  & \citep{gre030329} \\ 
 & 64.80 & $0.52 \pm 0.12$ & $12.55 \pm 4.63$ & $4.55\times10^{14}$ & VLT &  & \citep{gre030329} \\ 
 & 65.28 & $0.31 \pm 0.07$ & $24.50 \pm 6.94$ & $4.55\times10^{14}$ & VLT &  & \citep{gre030329} \\ 
 & 84.96 & $0.57 \pm 0.09$ & $53.85 \pm 4.08$ & $4.55\times10^{14}$ & VLT &  & \citep{gre030329} \\ 
 & 85.44 & $0.53 \pm 0.08$ & $57.08 \pm 4.06$ & $4.55\times10^{14}$ & VLT &  & \citep{gre030329} \\ 
 & 85.92 & $0.42 \pm 0.10$ & $62.21 \pm 6.10$ & $4.55\times10^{14}$ & VLT &  & \citep{gre030329} \\ 
 & 135.84 & $1.68 \pm 0.18$ & $66.32 \pm 3.38$ & $4.55\times10^{14}$ & NOT &  & \citep{gre030329} \\ 
 & 183.36 & $2.22 \pm 0.28$ & $75.16 \pm 3.32$ & $4.55\times10^{14}$ & VLT &  & \citep{gre030329} \\ 
 & 185.04 & $< 1$ ($3\sigma$) & &  $8.4\times10^{9}$ & VLBA & & \citep{radio030329} \\
 & 230.16 & $1.33 \pm 0.14$ & $70.91 \pm 3.31$ & $4.55\times10^{14}$ & VLT &  & \citep{gre030329} \\ 
 & 326.40 & $2.04 \pm 0.57$ & $1.16 \pm 7.64$ & $4.55\times10^{14}$ & VLT &  & \citep{gre030329} \\ 
 & 540.00 & $0.58 \pm 0.10$ & $42.70 \pm 9.26$ & $4.55\times10^{14}$ & VLT &  & \citep{gre030329} \\ 
 & 696.00 & $1.49 \pm 0.56$ & $99.71 \pm 6.60$ & $4.55\times10^{14}$ & VLT &  & \citep{gre030329} \\ 
 & 900.00 & $1.48 \pm 0.48$ & $25.42 \pm 9.41$ & $4.55\times10^{14}$ & VLT &  & \citep{gre030329} \\ 
 & 1992 & $< 1.8$ ($3\sigma$) & &  $8.4\times10^{9}$ & VLBA & & \citep{radio030329bis} \\
 & 5208 & $< 4.7$ ($3\sigma$) & &  $8.4\times10^{9}$ & VLBA & & \citep{radio030329bis} \\
\hline
\end{tabular}
\end{center}
\end{footnotesize}
\label{tab:030329pol}
\end{table*}

The observations of GRB\,030329 signaled, symbolically, the end of the ``golden age" of GRB afterglow polarimetry.  In fact, it became clear that the phenomenology offered by the GRBs even during the afterglow was much richer than expected, that models required a large set of possible and weakly constrained additions, and that in any case a full coverage of the polarimetric time-evolution of a typical afterglow was a very demanding task even for a 8\,m class telescope.

While activities for the late afterglow continued, the attention of the community was more and more devoted to the early afterglow and the reverse shock phases that were becoming observationally accessible thanks to a new generation of intermediate size robotic telescopes and detectors. The first result came for GRB\,060418, an event that was covered by robotic telescopes in the optical starting from a minute by the high-energy alert \citep{mol060418} and in polarimetry by the Liverpool Telescope\footnote{http://telescope.livjm.ac.uk} (LT) after about three minutes, in a large band roughly centered on the $V$ and $Rc$ filters \citep{mund060418}. The observations yielded only a mild upper limit, $P < 8$\%, that however was of considerable importance since it was obtained in a previously unexplored region of the GRB afterglow emission. Depending on the energy content of the outflow, e.g. a hot fireball or a Poynting flux dominated scenario \citep{covsci}, it is possible to observe a rather large linear polarization at early time in particular if the optical emission comes from the reverse-shock or there is large scale ordered magnetic field \citep{granot03b,rossipol,laz020813,sagivplasma}. A detailed analysis of the early-time light curve observed by the REM\footnote{http://www.rem.inaf.it} telescope \citep{mol060418} shows that very likely the initial optical peak of GRB\,060418 is due to the forward-shock onset, and that reverse-shock emission for this event was at most comparable to the one from the forward-shock \citep{mund060418,jinfan060418}, thus diluting the total polarization below the derived limit. Therefore large-scale ordered magnetic fields are not dominant in the afterglow emission of GRB\,060418 at early times. 

VLT observations close to an achromatic break, likely a jet break, in the afterglow light-curve were obtained for GRB\,071010A \citep{cov071010}, yielding a strong upper limit at $P < 1.3$\%. Single epoch measurements do not intrinsically allow an unambiguous interpretation, yet such a low value might argue against a structured jet since close to the jet-break time it is expected to record the maximum linear polarization for a given event. Spectropolarimetry with the VLT was obtained for GRB\,080928, showing a mild polarization level at about 2.5\%. VLT observations of the GRB\,080330 afterglow \citep{littlejohns080310} resulted in upper limits at the 2-3\% level, and therefore not able to put specific constraints on the afterglow modeling and the jet structure.

A breakthrough occurred with the early-time observations of GRB\,090102 \citep{ste090102} with the LT telescope. A high linear polarization at $\sim 10$\% level was measured about 3\,minutes after the high energy event. The analysis of the optical light-curve of GRB\,090102 \citep{gen090102} suggested that the early-time emission could be interpreted as the decaying part of a reverse-shock. The simplest interpretation for the high polarization is that a large-scale ordered magnetic field is driving the relativistic outflow, and this would be the first direct evidence of such magnetic fields in these sources. A high (and declining with time) polarization during the reverse-shock phase is a common feature of magnetic models of GRBs \citep[e.g., ][]{icmart}. 

\begin{table*}
\caption{Circular polarization measurements carried out for several GRB afterglows.}
\begin{footnotesize}
\begin{center}
\begin{tabular}{lccccll}
\hline
\hline
Event & $t-t_0$ & P$_{\rm circ}$ & $\nu$ & Instrument & z & Ref \\
          & (hour)  & (\%)             & (Hz)  \\
\hline
GRB\,990123 & 30.0  & $< 32$ ($3\sigma$) &  $8.46\times10^9$ & VLA & 1.6 & \citep{kul990123} \\
			&		&					&				&	&	& \citep{radiopol} \\
GRB\,991216 & 35.8 &  $< 17$ ($3\sigma$)  &  $8.46\times10^9$ & VLA & 1.02 & \citep{radiopol} \\
		      & 64.3 &  $< 15$ ($3\sigma$)  &  $8.46\times10^9$ & VLA & & \citep{radiopol} \\  
GRB\,020405 & 28.6 & $< 19$ ($3\sigma$)  &  $8.46\times10^9$ & VLA & 0.69 & \citep{radiopol} \\
GRB\,091018 & 3.74 & $< 0.15$ ($2\sigma$) & $4.55\times10^{14}$ & VLT & 0.97 & \citep{wier091018} \\
GRB\,121024A & 3.59 & $0.61 \pm 0.13$ & $4.55\times10^{14}$ & VLT & 2.30 & \citep{wier121024} \\
GRB\,130427 	& 36 &  $< 2.7$ ($3\sigma$) & $4.8\times10^{9}$ & EVN & 0.34 & \citep{horst130427} \\
			& 60 &  $< 5.7$ ($3\sigma$) & $48\times10^{9}$ & EVN &  &   \citep{horst130427} \\
			& 110 &  $< 15$ ($3\sigma$) & $48\times10^{9}$ & EVN &  &   \citep{horst130427} \\
\hline
\end{tabular}
\end{center}
\end{footnotesize}
\label{tab:cirpol}
\end{table*}

\begin{figure*}
\begin{center}
\includegraphics[width=14cm]{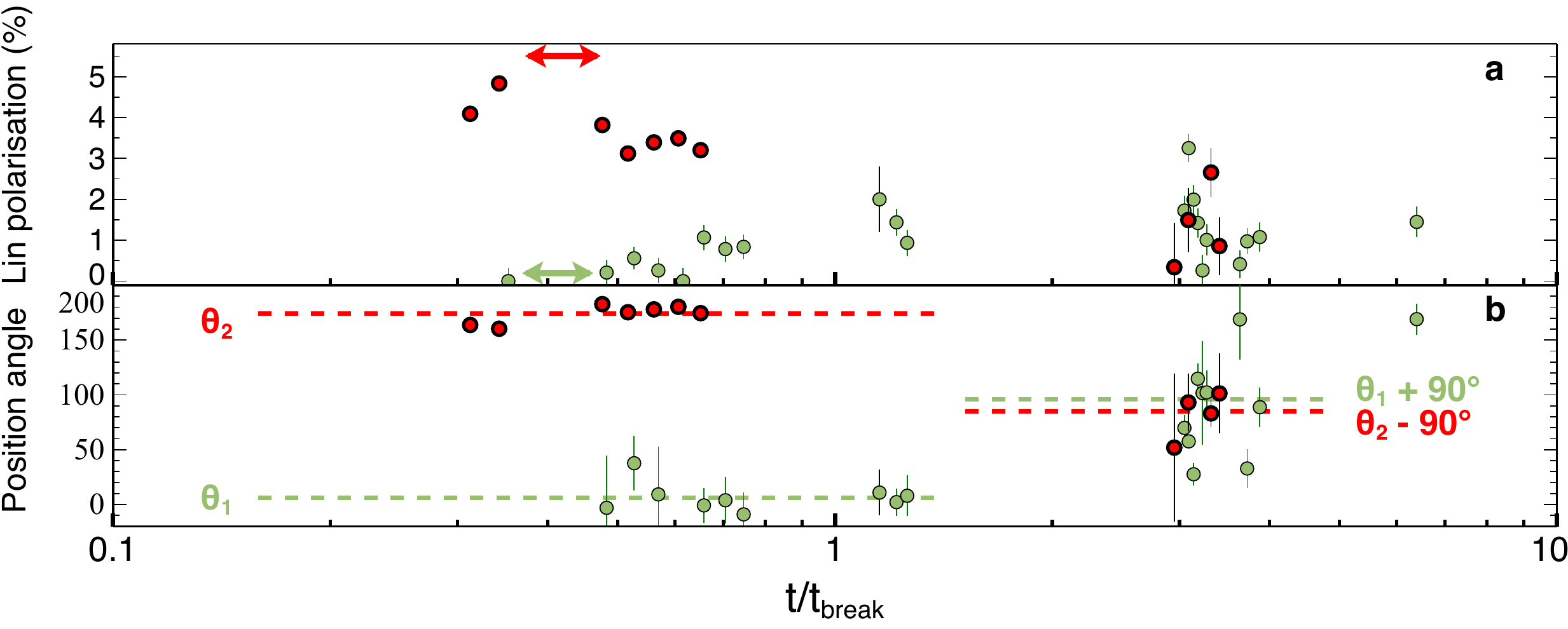}
\caption{Linear polarization (top) and polarization angle (bottom) as a function of the time since burst in terms of the break time in the optical light-curves of GRB\,091018 (green) and GRB\,121014A (red). The dotted lines show the average position angle before and after the jet-break to show the predicted 90$^\circ$ position angle swing. From \citet{wier121024}.}
\label{fig:polangrot}
\end{center}
\end{figure*}

Another large dataset was obtained with the ESO-VLT for GRB\,091018 \citep{wier091018}, comprising optical linear polarimetry, covering the evolution of the afterglow within 0.13 - 2.3\,days after the burst, and deep optical circular polarimetry. Near-infrared linear polarimetry was also obtained. The afterglow evolution was also very well sampled allowing the authors a very reliable analysis of the polarization evolution. The linear polarization degree shows variability from 0 up to 3\% both on short and long time-scales. For the circular polarization an upper limit of $P_{\rm circ} < 0.15$\% was derived. Linear polarization data are reported in Table\,\ref{tab:091018pol} while circular polarization data are reported in Table\,\ref{tab:cirpol}. The analyses of the light-curve allowed the identification of an achromatic break, the so-called jet-break, and the polarimetric observations well cover the earlier and later evolution of the afterglow. The initial part shows a smooth increase up to $P_{\rm lin} \sim 2$\% rather well described by a model assuming a homogenous jet \citep{rossipol}. After the jet-break the polarimetric behavior is more chaotic, possibly due to the presence of low-intensity bumps in the total light-curve. Nevertheless, the position angle seems to show a rotation by $\sim 90^\circ$, as predicted by the geometric models \citep{ghisellini99,sarippm}. This is the first possible identification of the position angle swing in a GRB afterglow polarization curve (Fig.\,\ref{fig:polangrot}). The low level of circular polarization detected during the first hours of afterglow evolution is also in agreement with the expectations for an afterglow not characterized by a strong ordered magnetic field \citep{tomaradiopol}. 

\begin{table*}
\caption{Linear polarization measurements carried out for the afterglows of GRB\,091018.}
\begin{footnotesize}
\begin{center}
\begin{tabular}{lcccccll}
\hline
\hline
Event & $t-t_0$ & P$_{\rm lin}$ & $\Theta$ & $\nu$ & Instrument & z & Ref \\
          & (hour)  & (\%)       & (deg)       & (Hz)  \\
\hline
GRB\,091018 & 3.17  & $< 0.32$ ($1\sigma$) &  & $4.55\times10^{14}$ & VLT & 0.97 & \citep{wier091018} \\ 
		& 4.33 & $0.21 \pm 0.31$ & $177.0 \pm 47.5$ & $4.55\times10^{14}$ & VLT &  & \citep{wier091018} \\
		&  4.73 & $0.56 \pm 0.27$ & $37.7 \pm 24.7$ & $4.55\times10^{14}$ & VLT &  & \citep{wier091018} \\
		& 5.11 & $0.26 \pm 0.31$ & $9.2 \pm 43.7$ & $4.55\times10^{14}$ & VLT &  & \citep{wier091018} \\
		& 5.53 & $< 0.32$\footnotetext{$1\sigma$}  &  & $4.55\times10^{14}$ & VLT &  & \citep{wier091018} \\
		&  5.91 & $1.07 \pm 0.30$ & $179.2 \pm 16.1$ & $4.55\times10^{14}$ & VLT &  & \citep{wier091018} \\
		& 6.33 & $0.78 \pm 0.31$ & $3.9 \pm 21.2$ & $4.55\times10^{14}$ & VLT &  & \citep{wier091018} \\
		& 6.70 & $0.84 \pm 0.30$ & $171.1 \pm 19.9$ & $4.55\times10^{14}$ & VLT &  & \citep{wier091018} \\
		& 10.34 & $2.0 \pm 0.8$ & $10.9 \pm 20.9$ & $1.39\times10^{14}$ & VLT &  & \citep{wier091018} \\		
		& 10.92 & $1.44 \pm 0.32$ & $2.2 \pm 12.6$ & $4.55\times10^{14}$ & VLT &  & \citep{wier091018} \\
		& 11.30 & $0.94 \pm 0.32$ & $8.0 \pm 18.6$ & $4.55\times10^{14}$ & VLT &  & \citep{wier091018} \\
		& 27.35 & $1.73 \pm 0.36$ & $69.8 \pm 11.7$ & $4.55\times10^{14}$ & VLT &  & \citep{wier091018} \\
		& 27.73 & $3.25 \pm 0.35$ & $57.6 \pm 6.1$ & $4.55\times10^{14}$ & VLT &  & \citep{wier091018} \\
		& 28.16 & $1.99 \pm 0.35$ & $27.6 \pm 10.0$ & $4.55\times10^{14}$ & VLT &  & \citep{wier091018} \\
		& 28.54 & $1.42 \pm 0.36$ & $114.6 \pm 14.0$ & $4.55\times10^{14}$ & VLT &  & \citep{wier091018} \\
		& 28.97 & $0.27 \pm 0.38$ & $101.6 \pm 47.1$ & $4.55\times10^{14}$ & VLT &  & \citep{wier091018} \\
		& 29.35 & $1.00 \pm 0.38$ & $102.2 \pm 20.2$ & $4.55\times10^{14}$ & VLT &  & \citep{wier091018} \\
		& 32.64 & $0.41 \pm 0.34$ & $168.8 \pm 36.6$ & $4.55\times10^{14}$ & VLT &  & \citep{wier091018} \\
		& 33.40 & $0.97 \pm 0.32$ & $32.8 \pm 17.8$ & $4.55\times10^{14}$ & VLT &  & \citep{wier091018} \\
		& 34.78 & $1.08 \pm 0.35$ & $88.7 \pm 17.9$ & $4.55\times10^{14}$ & VLT &  & \citep{wier091018} \\
		& 57.37 & $1.45 \pm 0.37$ & $169.0 \pm 14.3$ & $4.55\times10^{14}$ & VLT &  & \citep{wier091018} \\
\hline
\end{tabular}
\end{center}
\end{footnotesize}
\label{tab:091018pol}
\end{table*}

One more measurement of high polarization at early times was carried out for GRB\,091208B with the Kanata 1.5m telescope\footnote{http://hasc.hiroshima-u.ac.jp/telescope/kanatatel-e.html} obtaining $P \sim 10$\% several minutes after the events \citep{ue091208}. The early-time afterglow light-curve is consistent with a standard forward-shock emission. The high symmetry of the early afterglow evolution phase should be characterized by a low polarization level, if any, unless some other component is present, i.e. a reverse shock, or the symmetry is broken as due to a large scale magnetic field. If the emission is indeed due to the forward-shock the magnetic field is not likely advected from the central source by the expanding outflow and should be generated locally. \citet{ue091208} suggest that if the shock sweeps inhomogeneous external medium instabilities can grow producing strong random magnetic fields on large scales \citep{sirgood,inouermi} that could decay slow enough to survive in the entire emission region. Applying the same idea developed for the ``patchy-model" by \citet{patchy}, the length scale of these fluctuations must be $l_p \sim 4 \times 10^{14}$\,cm and the polarization angle should change randomly with time. 

The interest in the early-time GRB optical polarization is also shown by the development of instruments able to measure linear polarimetry of very bright GRB optical counterparts as the MASTER Global Robotic Net\footnote{http://observ.pereplet.ru}. Some of these measurements could possibly be more related to the prompt emission rather than the afterglows. Yet we report them here for completeness. In \citet{gorbmaster} polarization at several percent for GRB\,091127 was reported but a reanalysis of the data showed that it was likely an artifact due to adverse atmospheric conditions \citep{gorbmasterlarge}. Upper limits at about 10\% and 15\% are also reported for GRB\,100906A and GRB\,121011A during the first hour of afterglow evolution \citep{gorb100906,prumaster}. Sparse data indicating a possible polarization at a few per cent level during early afterglow evolution of GRB\,110205A were also reported by \citet{cucc2011} and \citet{goros110205} with the LT and the CAHA. 

An amazing discovery, made again with the LT, was the observation of a decaying polarization, with an essentially constant position angle, with a maximum polarization at about $30$\% several minutes after the high-energy event of GRB\,120308A \citep{mundell13}. The constancy of the position angle basically rules out plasma or magnetohydrodynamical instabilities that are not supposed to show coherent properties during the early afterglow evolution. Modeling in addition the early-time light-curve with contributions both from the reverse- and forward-shock, these observations imply a magnetized baryonic jet with a large-scale uniform field \citep{mundell13,lyupol}. GRB\,120308A polarimetry was also analyzed in detail by \citet{zhajin} using all the available afterglow data. They derived that the strength of the magnetic field in the reverse-shock region should have been an order of magnitude stronger than in the forward-shock region. As a consequence, the outflow turned out to be mildly magnetized, at a level $\sigma$ of a few percent. The polarimetric observations therefore definitely show that for at least some GRBs a relevant fraction of the energy is released in the form of Poynting flux. \citet{lanearly} analyzed the total and polarized early-time curves for this event deriving that both a toroidal and aligned with the jet axis configurations for the magnetic field are possible. 

Relatively later time observations cannot follow the diagnostically important early afterglow but can carry out observations with bigger telescopes allowing us to test unexplored regimes. This was the case of GRB\,121024A that was intensively observed with the VLT \citep{wier121024} starting from a few hours after the GRB and obtaining a positive and highly  unexpected detection of circular polarization, $P_{\rm circ} = 0.61\ \pm 0.13$\%, together with an extensive linear polarimetric monitoring (Table\,\ref{tab:121024pol}). The first observations showed a rather high polarization level, $P_{\rm lin} \sim 5$\%, with a global decreasing trend with time and a constant position angle. Observations carried out the night after, showed a lower polarization level with a clear $90^\circ$ rotation of the position angle. Analysis of the light-curve allowed us to identify a jet-break between the two sets of observations and this is a very clear identification of the polarization angle swing predicted to occur around the jet-break time of a homogeneous jet that is not spreading sideways \citep{rossipol}. During the circular polarimetry measurement the linear polarization was about 4\%, and therefore the circular to linear polarimetry ratio turned out to be $P_{\rm circ}/P_{\rm lin} \sim 0.15$, a very high value, orders of magnitudes greater than the theoretical expectations \citep{matiocirc,sagivplasma,tomaradiopol}. If the emission process is synchrotron the expected polarization is indeed $P_{\rm circ} \sim \gamma_{\rm e}^{-1}$, where $\gamma_{\rm e}$ is the random Lorentz factor of the accelerated electrons emitting the observed radiation. This relation holds under the assumption of isotropic pitch-angle distribution and ordered magnetic fields \citep{tomaradiopol}, and the high value of measured circular polarization poses a challenge to this assumption. Furthermore, a detailed analysis carried out by \citet{navacirc} suggests that under the hypothesis of optically thin synchrotron emission such a high value of circular to linear polarization ratio is not possible even with extremely isotropic pitch-angle distribution. A satisfactory interpretation of this striking result is still missing.

\begin{table*}
\caption{Linear polarization measurements carried out for the afterglows of GRB\,121024A.}
\begin{footnotesize}
\begin{center}
\begin{tabular}{lcccccll}
\hline
\hline
Event & $t-t_0$ & P$_{\rm lin}$ & $\Theta$ & $\nu$ & Instrument & z & Ref \\
          & (hour)  & (\%)       & (deg)       & (Hz)  \\
\hline
GRB\,121024A & 2.69  & $4.09 \pm 0.20$ & $163.7 \pm 2.8$  & $4.55\times10^{14}$ & VLT & 2.30 & \citep{wier121024} \\ 
			& 2.96 & $4.83 \pm 0.20$ & $160.3 \pm 2.3$ & $4.55\times10^{14}$ & VLT &  & \citep{wier121024} \\
			& 4.11 & $3.82\pm 0.20$ & $182.7 \pm 3.0$ & $4.55\times10^{14}$ & VLT &  & \citep{wier121024} \\
			& 4.46 & $3.12 \pm 0.19$ & $175.3 \pm 3.5$ & $4.55\times10^{14}$ & VLT &  & \citep{wier121024} \\
			& 4.84 & $3.39 \pm 0.18$ & $178.0 \pm 2.9$ & $4.55\times10^{14}$ & VLT &  & \citep{wier121024} \\
			& 5.23 & $3.49 \pm 0.18$ & $180.3 \pm 3.0$ & $4.55\times10^{14}$ & VLT &  & \citep{wier121024} \\
			& 5.62 & $3.20 \pm 0.18$ & $174.5 \pm 3.3$ & $4.55\times10^{14}$ & VLT &  & \citep{wier121024} \\
			& 25.45 & $0.34 \pm 1.09$ & $51.9 \pm 67.5$ & $4.55\times10^{14}$ & VLT &  & \citep{wier121024} \\
			& 26.62 & $1.49 \pm 0.78$ & $93.1 \pm 26.6$ & $4.55\times10^{14}$ & VLT &  & \citep{wier121024} \\
			& 28.62 & $2.66 \pm 0.60$ & $83.0 \pm 12.6$ & $4.55\times10^{14}$ & VLT &  & \citep{wier121024} \\
			& 29.39 & $0.86 \pm 0.72$ & $101.3 \pm 36.4$ & $4.55\times10^{14}$ & VLT &  & \citep{wier121024} \\
\hline
\end{tabular}
\end{center}
\end{footnotesize}
\label{tab:121024pol}
\end{table*}

Radio observations at 4.8\,GHz with the EVN\footnote{http://www.evlbi.org} reporting upper limits at a few per cent levels both for linear and circular polarimetry of GRB\,13042A7 were obtained by \citet{horst130427}, while a low-significance polarization measurement obtained with the Skinakas Observatory 1.3\,m telescope\footnote{http://skinakas.physics.uoc.gr/en/}, during the first two hours after GRB\,131003A, are reported by \citet{king131030}. The LT obtained a lower limit at about 22\% for GRB\,140430A when the prompt phase was still active, a few minutes after the high-energy event \citep{kopac15}. How to interpret the limits depends on how many components were contributing to the measured optical photons, i.e. reverse- and forward-shock and in which proportion. The analysis of the early-time observations for this event unfortunately did not allow us to derive a firm conclusion. Finally, the MASTER network was able to obtain a rather interesting lower limit, at about 8\%, 1-2 minutes after GRB\,150301B \citep{gorbovskoy16}.

\section{Polarimetry and Lorentz invariance violation}


\label{sec:liv}
An interesting aspect related to measurements of linear polarization of GRBs and in general of cosmological sources is the possibility to constrain Lorentz invariance violations (LIV), i.e. the invariance of the laws of physics under rotation and boosts. 

In general, it is possible to describe light as composed of two independently propagating constituent waves, each possessing a polarization and a velocity. Certain forms of relativity violations cause light to experience birefringence, a change in polarization as it propagates. The changes grow linearly with distance travelled, so birefringence over cosmological scales offers a sensitive probe for relativity violations. Searches for this vacuum birefringence using polarized light emitted from sources at cosmological distances yield some of the sharpest existing tests of relativity \citep{firstliv}. In addition, polarimetry of a large number of cosmological sources can also allow interesting tests for the existence and physical properties of very light axion-like particles \citep{bassan,menaxions}.

The possible unification at the Planck energy scale of the theory of General Relativity and the quantum theory in the form of the Standard Model requires to quantize gravity, which can lead to fundamental difficulties: one of these is to admit the Lorentz Invariance Violation (LIV)
\citep[e.g.][]{jacobson06,liberati09,mattingly05}

A possible experimental test for such violation is to measure the helicity dependence
of the propagation velocity of photons \citep[see e.g.][and references
therein]{laurent11a}. The light dispersion relation is given in this
case by

\begin{equation}
\omega^{2}=k^{2}\pm\frac{2\xi k^{3}}{M_{Pl}}\equiv\omega^{2}_{\pm}
\label{eq:dispersion1}
\end{equation}

where $E=\hbar\omega$, $p=\hbar k$, $M_{Pl}$ is the Planck Mass, and the sign of the cubic term is determined by the chirality (or circular polarization) of the photons, which leads to a rotation of the polarization during the propagation
of linearly polarized photons. This effect is known as vacuum birefringence.

Equation \ref{eq:dispersion1} can be approximated as follows

\begin{equation}
\omega_{\pm}=\vert k \vert \sqrt{1\pm\frac{2\xi k}{M_{Pl}}}\approx\vert k\vert(1\pm\frac{\xi k }{M_{Pl}})
\label{eq:dispersion2}
\end{equation} 

where $\xi$ gives the order of magnitude of the effect. In practice some quantum-gravity theories \citep[e.g.][]{myers03} predict that the polarization plane of the electromagnetic waves emitted by a distant source rotates by a quantity $\Delta\theta$ while the latter propagates through space, and this as a function of the energy of the photons, see Eq. \ref{eq:rotation}, where $d$ is the distance of the source:

\begin{equation}
\Delta\theta(p)=\frac{\omega_{+}(k)-\omega_{-}(k)}{2}\ d\approx\xi\frac{k^{2}d}{2M_{Pl}}
\label{eq:rotation}
\end{equation}

As a consequence the signal produced by a linearly polarized source, observed in a given energy band could vanish, if the distance is large enough, since the differential rotation acting on the polarization angle as a function of energy would in the end add opposite oriented polarization vectors, and hence in a net un-polarized signal.
But this effect is very tiny, since it is inversely proportional to the Planck Mass ($M_{Pl}\sim$2.4$\times$10$^{18}$ GeV), the observed source needs to be at cosmological distances. The simple fact to detect the polarization signal from a distant source, can put a limit to such a possible violation. This experiment has been performed recently by \citet{laurent11a}, \citet{toma12}, and \citet{gotz13} making use of the prompt emission of GRBs. 
Indeed, since GRBs are at the same time at cosmological distances, and emitting at high energies, their polarization measurements are highly suited to measure and improve upon these limits.
\citet{laurent11a}, taking advantage of the polarization measurements obtained with IBIS on GRB 041219A in different energy bands (200--250 keV, 250--325 keV), and from the measure of distance of the source (z$>$0.02 at 90\% c.l., equivalent to a luminosity distance 85 Mpc) were able to set the most stringent limit to date to a possible LIV effect: $\xi <$1.1$\times$10$^{-14}$. We note that, although \citet{toma12} claim to have derived a more stringent limit ($\xi <$8$\times$10$^{-16}$), their measure does not rely on a real measure of the distance of the GRBs they analyse, but they use a distance estimate based on an empirical spectral-luminosity relation \citep{yonetoku10}, whose selection effects, physical interpretation, and absolute calibration are not yet completely understood.
By using the distance measured from the afterglow absorption spectrum of GRB140206A (23 Gpc) \citet{gotz14} obtained

\begin{equation}
\xi < \frac{2 M_{Pl}\Delta\theta(k)}{(k_{2}^{2}-k_{1}^{2})\ d}\approx 1\times 10^{-16},
\label{eq:xi}
\end{equation}

improving the previous limit obtained by the same authors on GRB 061122 \citep{gotz13} by a factor of three.

Another powerful LIV test was carried out by \citet{fanliv} by means of the spectopolarimetric observations of the optical afterglows of GRB\,020813 and GRB\,021004. 
Since linear polarization is a superposition of two monochromatic waves with opposite circular polarizations (see Eq. \ref{eq:dispersion2}), the plane of linear polarization is subject to a rotation along the photons' path because of the difference between the two circular components. For a photon of frequency $\nu_{\rm obs}$ emitted at redshift $z$ and with intrinsic polarization $\Phi_0$ we have\footnote{The original Eq. 3 in \citep{fanliv} contained a typographical error here corrected.}:
\begin{equation}
\Phi_{(n)} = \Phi_0 + 7.8 \times 10^{60} \xi \frac{l^{n+1}_p}{c^{n+1}} (2\pi\nu_{\rm obs})^{n+1} F(z, n),
\end{equation}
where $l_p = \sqrt{\hbar G / c^3} = \hbar c / E_{\rm pl}$ is the Planck's length-scale, $c$ the speed of light, $G$ the gravitational constant, and $F(z,n)$ a function that depends on the adopted cosmology. For the concordance cosmology, $F(z,n) \simeq 1$ at redshift $z\sim1$, and increases only slowly at higher redshift. On the other hand, the dependence on the photon energy is stronger, $\nu^{n+1}$, immediately showing the importance of obtaining polarimetry of cosmological sources at the highest possible energies.

\citet{fanliv} analyzed the case with $n=1$ and, by the lack of any rotation of the polarization plane of the spectra of GRB\,020813 and GRB\,021004, they could constrain $| \xi | < 2 \times 10^{-7}$ at $3\sigma$.

We mention in passing that other limits on cosmological birefringence were obtained by means of polarization studies of different classes of cosmological sources, e.g. radio-galaxies \citep{disereradiogal}.

\section{Conclusions}
\label{sec:conclusions}


In this paper we have separated the prompt and afterglow GRB phases mainly for making the presentation easier to follow and because the observational techniques and, to some extent, the theoretical scenarios are different. Nevertheless, some of the general conclusions hold for both phases and the available observational material definitely provides one of the most relevant set of constraints to the large family of models and parameters describing the GRB phenomenologies.


The large set of observations available for the afterglows, mainly but not only in the optical (Section\,\ref{sec:aftobs}), allows us to derive a few important conclusions. First of all, the simple observations of variable polarization implies that the afterglow radiation is intrinsically polarized, thus offering strong observational evidence for the synchrotron origin of the afterglow emission, although different scenarios cannot yet be completely ruled out.
The observations of specific patterns (i.e. the position angle swing) during the evolution of the afterglows in polarimetry that have been predicted in advance give also confidence to the general interpretative scenario, although exceptions are present. And the detection of circular polarimetry at a level much higher than expected instead poses a formidable challenge to our present GRB afterglow emission interpretation. The solid observations of high polarization during the early-afterglow strongly implies that at least some of the GRBs have an important magnetic energy content. A striking discovery in itself.

The success of recent observational campaigns clearly show that a massive approach, trying to follow the afterglow evolution from the early-time, with intermediate-size robotic telescope, to the late phases, with the biggest available facilities, is required. And the parameter space for discoveries is still huge. Radio observations are promising, in particular with future high-sensitivity facilities, and mm observations with ALMA can help to dramatically extend the energy range of the observations and the testing capabilities of the various interpretative scenarios.

For the prompt phase the situation is less clear, but also offering perspective for exciting discoveries in the near future. A final answer to distinguish between intrinsic and geometric models could be obtained by accumulating more observations. Indeed, models (1-2, 6) -- as defined in Section \ref{sec:prompt_theory} -- predict a polarized emission for all bursts, whereas models (3-5) would predict that only a small fraction of GRBs are highly polarized. 
This shows the importance of accumulating polarimetric measurements for the
understanding of intrinsic properties of GRBs, but the current instrumentation is statistically limited and can provide measurements just for the brightest events. 

Nevertheless, although all currently available measures (see Table \ref{tab:polasummary}), taken individually, have not a very high significance ($\gtrsim$3 $\sigma$), they indicate that GRBs are indeed good candidates for highly $\gamma$-ray polarized sources, and that they are prime targets for future polarimetry experiments.
On the other hand, as can be seen from Table \ref{tab:polasummary} the currently available GRB sample does not show extreme spectral characteristics, e.g. in terms of peak energy, but they are on the upper end of the GRB fluence distribution. This means that, on one hand, this sample may be well representative of the whole GRB population. On the other hand the fluence bias is clearly an instrumental selection effect due to the high photon statistics needed to perform the polarization measurements in IBIS and GAP.

As discussed above, prompt polarization features can be explained by synchrotron radiation in an ordered magnetic field \citep{granot03a,granot03b,nakar03}, by the jet structure \citep{lazzati09}, or , independently from the magnetic field structure or the emission processes, by the observer's viewing angle with respect to the jet \citep{lazzati04}, even in the case of thermal radiation from the jet photosphere \citep{lundman14}. In addition the level of magnetization of the jet can also play a role \citep{spruit01,lyutikov06}. For instance the ICMART model \citep{icmart}, which implies a magnetically dominated wind launched by the central engine, predicts a decrease of the polarization level during GRB individual pulses, but this hypothesis cannot be tested with the current data. Indeed, as pointed out by \citet{toma09}, the different models are hardly distinguishable relaying only on $\gamma$-ray data, and a result can be achieved only on statistical grounds, i.e. having a sample of several tens of measures at high energies. This will hardly be achieved before the advent of dedicated GRB polarimetry experiments, e.g. POLAR \citep{polar} or POET \citep{poet}. 

Recently, a few polarization measurements of the very early optical afterglow have been reported sometime while the prompt high-energy phase was still on going (see also Section \ref{sec:aftobs}). While \citet{kopac15} and \citet{gorbovskoy16} do not report significant detections for GRB140430A and GRB150413A respectively, for GRB 150301B a lower limit of 8\% has been reported \citep{gorbovskoy16} (the earliest measurement in the co-moving time frame to date) and for GRB 120308A a high level, $\Pi$=28$\pm$4\%, of linear optical polarization in the early afterglow has been reported by \citet{mundell13}. The latter measure allowed us to point out the presence of a magnetized reverse shock with an ordered magnetic field, confirming the presence of high magnetic fields in the GRB ejecta, and indicating that the multi-wavelength approach could be fruitful, even if there is currently no consensus on the common origin of the $\gamma$-ray and optical emission in the prompt phase of GRBs \citep[e.g.][]{vestrand05,stratta09,gotz11,guidorzi11}.

\section*{Acknowledgments}

This work has been supported by ASI grant I/004/11/2. SC thanks Gabriele Ghisellini for invaluable, in number and quality, discussions and suggestions. Davide Lazzati for having shared the beginning of this quest long time ago. Finally, a special mention for Javier Gorosabel Urkia, a friend and colleague who left us too early. DG acknowledges the financial support of the UnivEarthS Labex program at Sorbonne Paris Cit\'e (ANR-10-LABX-0023 and ANR-11-IDEX-0005-02).

\bibliography{covinogotz_AT_v5}
\bibliographystyle{aa}


%


%

\end{document}